\documentclass[twocolumn]{aastex631}

\usepackage{hyperref} 
\usepackage{longfigure}
\usepackage{amsmath}
\usepackage[version=4]{mhchem} 
\usepackage{multirow}
\usepackage{booktabs}

\DeclareRobustCommand{\ion}[2]{%
\relax\ifmmode
\ifx\testbx\f@series
{\mathbf{#1\,\mathsc{#2}}}\else
{\mathrm{#1\,\mathsc{#2}}}\fi
\else\textup{#1\,{\mdseries\textsc{#2}}}%
\fi}

\defcitealias{Acebron2022}{A22}
\def\GLEE{\textsc{Glee}}

\shorttitle{Extended Strong Lensing Model of SDSS~J1029$+$2623}
\shortauthors{Acebron et al.}
\graphicspath{{./}{figures/}}

\begin{document}


\title{The next step in galaxy cluster strong lensing:\\modeling the surface brightness of multiply-imaged sources \footnote{This work is based in large part on data collected at ESO VLT (prog. ID 0102.A-0642(A)) and NASA \textit{HST}.}}

\correspondingauthor{Ana Acebron}
\email{ana.acebron@unican.es}

\author[0000-0003-3108-9039]{Ana Acebron}
\affiliation{Dipartimento di Fisica, Universit\`a degli Studi di Milano, Via Celoria 16, I-20133 Milano, Italy}
\affiliation{INAF -- IASF Milano, via A. Corti 12, I-20133 Milano, Italy}
\affiliation{Instituto de Física de Cantabria (CSIC-UC), Avda. Los Castros s/n, 39005 Santander, Spain}

\author[0000-0002-5926-7143]{Claudio Grillo}
\affiliation{Dipartimento di Fisica, Universit\`a degli Studi di Milano, Via Celoria 16, I-20133 Milano, Italy}
\affiliation{INAF -- IASF Milano, via A. Corti 12, I-20133 Milano, Italy}

\author[0000-0001-5568-6052]{Sherry H.~Suyu}
\affiliation{Technical University of Munich, TUM School of Natural Sciences, Physics Department, James-Franck-Str.~1, 85748 Garching, Germany}
\affiliation{Max-Planck-Institut f\"ur Astrophysik, Karl-Schwarzschild-Str.~1, D-85748 Garching, Germany}
\affiliation{Academia Sinica Institute of Astronomy and Astrophysics (ASIAA), 11F of ASMAB, No.1, Section 4, Roosevelt Road, Taipei 106216, Taiwan}

\author[0000-0003-1383-9414]{Giuseppe Angora}
\affiliation{INAF -- Osservatorio Astronomico di Capodimonte, Salita Moiariello 16, I-80131 Napoli, Italy}
\affiliation{INFN -- Sezione di Ferrara, via Saragat 1, I--44122, Ferrara, Italy}

\author[0000-0003-1383-9414]{Pietro Bergamini}
\affiliation{Dipartimento di Fisica, Universit\`a degli Studi di Milano, Via Celoria 16, I-20133 Milano, Italy}
\affiliation{INAF -- OAS, Osservatorio di Astrofisica e Scienza dello Spazio di Bologna, via Gobetti 93/3, I-40129 Bologna, Italy}

\author[0000-0001-6052-3274]{Gabriel B. Caminha}
\affiliation{Technical University of Munich, TUM School of Natural Sciences, Physics Department, James-Franck-Str.~1, 85748 Garching, Germany}
\affiliation{Max-Planck-Institut f\"ur Astrophysik, Karl-Schwarzschild-Str.~1, D-85748 Garching, Germany}

\author[0000-0002-5085-2143]{Sebastian Ertl}
\affiliation{Max-Planck-Institut f\"ur Astrophysik, Karl-Schwarzschild-Str.~1, D-85748 Garching, Germany}
\affiliation{Technical University of Munich, TUM School of Natural Sciences, Physics Department, James-Franck-Str.~1, 85748 Garching, Germany}

\author[0000-0001-9261-7849]{Amata Mercurio}
\affiliation{Università di Salerno, Dipartimento di Fisica ``E.R. Caianiello", Via Giovanni Paolo II 132, I-84084 Fisciano (SA), Italy}
\affiliation{INAF -- Osservatorio Astronomico di Capodimonte, Salita Moiariello 16, I-80131 Napoli, Italy}
\affiliation{INFN – Gruppo Collegato di Salerno - Sezione di Napoli, Dipartimento di Fisica ``E.R. Caianiello", Università di Salerno, via Giovanni Paolo II, 132 - I-84084 Fisciano (SA), Italy.}

\author[0000-0001-6342-9662]{Mario Nonino}
\affiliation{INAF -- Osservatorio Astronomico di Trieste, via G. B. Tiepolo 11, I-34131 Trieste, Italy}

\author[0000-0002-6813-0632]{Piero Rosati}
\affiliation{Dipartimento di Fisica e Scienze della Terra, Universit\`a degli Studi di Ferrara, via Saragat 1, I-44122 Ferrara, Italy}
\affiliation{INAF -- OAS, Osservatorio di Astrofisica e Scienza dello Spazio di Bologna, via Gobetti 93/3, I-40129 Bologna, Italy}

\author[0000-0002-1293-5503]{Han Wang}
\affiliation{Max-Planck-Institut f\"ur Astrophysik, Karl-Schwarzschild-Str.~1, D-85748 Garching, Germany}
\affiliation{Technical University of Munich, TUM School of Natural Sciences, Physics Department, James-Franck-Str.~1, 85748 Garching, Germany}

\author[0000-0001-5976-9728]{Andrea Bolamperti}
\affiliation{Dipartimento di Fisica ed Astronomia, Università degli Studi di Padova, Vicolo dell’Osservatorio 3, I-35122 Padova, Italy}
\affiliation{Istituto Nazionale di Astrofisica (INAF), Osservatorio di Padova, Vicolo dell'Osservatorio 5, I-35122 Padova, Italy}
\affiliation{European Southern Observatory, Karl-Schwarzschild-Str.~2, D-85748 Garching, Germany}

\author[0000-0003-1225-7084]{Massimo Meneghetti}
\affiliation{INAF -- OAS, Osservatorio di Astrofisica e Scienza dello Spazio di Bologna, via Gobetti 93/3, I-40129 Bologna, Italy}

\author[0000-0003-2497-6334]{Stefan Schuldt}
\affiliation{Dipartimento di Fisica, Universit\`a degli Studi di Milano, Via Celoria 16, I-20133 Milano, Italy}
\affiliation{INAF -- IASF Milano, via A. Corti 12, I-20133 Milano, Italy}

\author[0000-0002-5057-135X]{Eros Vanzella}
\affiliation{INAF -- OAS, Osservatorio di Astrofisica e Scienza dello Spazio di Bologna, via Gobetti 93/3, I-40129 Bologna, Italy}





\begin{abstract}
Overcoming both modeling and computational challenges, we present, for the first time, the extended surface-brightness distribution model of a strongly-lensed source in a complex galaxy-cluster-scale system.
We exploit the high-resolution Hubble Space Telescope (HST) imaging and extensive Multi Unit Spectroscopic Explorer spectroscopy to build an extended strong-lensing model, in a full multi-plane formalism, of SDSS~J1029$+$2623, a lens cluster at $z~=~0.588$ with three multiple images of a background quasar ($z = 2.1992$). 
Going beyond typical cluster strong-lensing modeling techniques, we include as observables both the positions of 26 pointlike multiple images from seven background sources, spanning a wide redshift range between 1.02 and 5.06, and the extended surface-brightness distribution of the strongly-lensed quasar host galaxy, over $\sim78000$ HST pixels. In addition, we model the light distribution of seven objects, angularly close to the strongly-lensed quasar host, over $\sim9300$ HST pixels.
Our extended lens model reproduces well both the observed intensity and morphology of the quasar host galaxy in the HST F160W band (with a $0\arcsec.03$ pixel scale). The reconstructed source shows a single, compact, and smooth surface-brightness distribution, for which we estimate an intrinsic magnitude of 23.3 $\pm$ 0.1 in the F160W band and a half-light radius of (2.39 $\pm$ 0.03) kpc.
The increased number of observables enables the accurate determination of the total mass of line-of-sight halos lying angularly close to the extended arc. 
This work paves the way for a new generation of galaxy cluster strong-lens models, where additional, complementary lensing observables are directly incorporated as model constraints.
\end{abstract}




\section{Introduction} \label{sec:intro}
Until now, strong-lensing models of galaxy clusters have only considered the positions of \emph{pointlike} multiply-imaged sources as observables \citep[e.g.,][]{Lagattuta2019, Caminha2019, Acebron2020, Diego2020, Zitrin2020}.
The avenue of high-resolution, multi-band imaging of lens clusters, together with extensive ground- or space-based follow-up spectroscopy, has enabled important progress in the lens modeling of galaxy clusters. The secure identification of large samples of multiple images in their cores has resulted in a new generation of \emph{position-based} lens models, with an unprecedented level of precision and accuracy \cite[e.g.,][]{Bergamini2021, Richard2021, Bergamini2023, Caminha2023, Diego2023, Furtak2023, Cha2024}. 
In particular, multiply-lensed knots within resolved extended sources have been shown to efficiently constrain the position of the critical lines locally \cite[][]{Grillo2016, Bergamini2023, Lagattuta2023}, further proving that the accuracy and precision of cluster lens models, and therefore of any subsequent application, critically depend on the number of secure observables. 

The next logical step in the strong-lensing modeling of galaxy clusters is to increase the number of observables by directly modeling the extended surface-brightness distribution of multiply-imaged sources.
While the positions of \emph{pointlike} multiple images constrain the value of the deflection angle at their positions, the surface-brightness distribution of lensed sources provides information, in a complementary way, on the higher-order derivatives of the deflection angle, as well as probing the smaller magnification regime. 
Studies of this kind are key to fully exploiting the exquisite observations from current facilities, such as the Hubble Space Telescope (HST) and the JWST, but have mostly been carried out in galaxy- or group-scale lensing systems.

The large number of observables enables using such lenses as powerful astrophysical and cosmological tools.
For instance, it is possible to detect low-mass dark halos at cosmological distances through the induced distortions in spatially-extended lensed galaxies \citep{Vegetti2012, Despali2022, Nightingale2024} and probe alternative dark matter models \citep[e.g.,][]{Vegetti2018}. 
Extended strong-lensing models allow for robust and detailed studies of the inner mass distribution of the lenses, as well as the properties of the lensed background sources in galaxy- \citep{Grillo2008, Schuldt2019, Shajib2021, Ertl2023b},
and in group-scale systems \citep{Suyu2010, Wang2022, Bolamperti2023, Wang2024}.
The reconstruction of high-redshift lensed galaxies can also shed light on galaxy formation and evolution processes in the early Universe \citep{Chirivi2020, Rizzo2020}.  
Strongly-lensed, time-varying sources, such as quasars \citep[QSOs,][]{Schechter1997, Fassnacht2002, Fohlmeister2007, Courbin2011, Millon2020, Munoz2022} or supernovae \citep{Kelly2015, Goobar2017, Rodney2021, Pierel2023, Frye2024, Pierel2024a, Pierel2024b} represent alternative probes to measure the values of the cosmological parameters, in particular that of the present-day expansion rate of the Universe, set by the Hubble constant, $H_0$ \citep{Refsdal1964}.
By exploiting high-resolution imaging, with state-of-the-art lens modeling techniques \citep[see e.g.,][]{Suyu2010, Birrer2018}, that include spatially extended emissions of the QSO host galaxy in addition to the \emph{pointlike} QSO multiple image positions, robust measurements of the Fermat potential of the lens and thus of the value of $H_0$ have been obtained \citep{Suyu2017, Chen2019, Birrer2019, Wong2020, Shajib2023}. 
In addition, the modeling of strongly-lensed QSO host galaxies represents a unique way to investigate the co-evolution of supermassive black holes and their host galaxies beyond the local Universe \citep{Peng2006, Ding2017b}. The lensing magnification stretches and resolves the host galaxies out from the bright point-source emission, allowing for high-fidelity measurements of their total luminosities \citep{Ding2017a}.

Extending these analyses to lens galaxy clusters is no easy feat. The main challenges arise from the increased level of complexity in modeling their total mass distribution and the larger modeling areas involved. Indeed, while the larger number of observables provides more information compared to galaxy-scale analyses, it comes at the cost of a significant increase in computational times and Random Access Memory (RAM) allocations. 
Surface-brightness models of multiply-lensed sources in cluster-scale systems have been exploited to reduce the intrinsic degeneracy between the truncation radius and central velocity dispersion parameters of cluster galaxies (lying close in projection to extended images) as well as studying their internal mass structure \citep{Grillo2008, Monna2015, Monna2017, Galan2024}. 
However, contrarily to these analyses, where the mass parameters associated with the cluster total mass distribution were kept fixed or approximated by an external shear component in the optimization, we aim at developing, for the first time, a fully extended strong-lensing model of a strongly-lensed source in a galaxy cluster. Such improved modeling techniques of lens galaxy clusters enable robust studies of the intrinsic properties of high-redshift, strongly-lensed sources as well as cosmological applications \citep{Xie2024}.

\begin{figure*}[ht]
\centering
\includegraphics[width=0.8\linewidth]{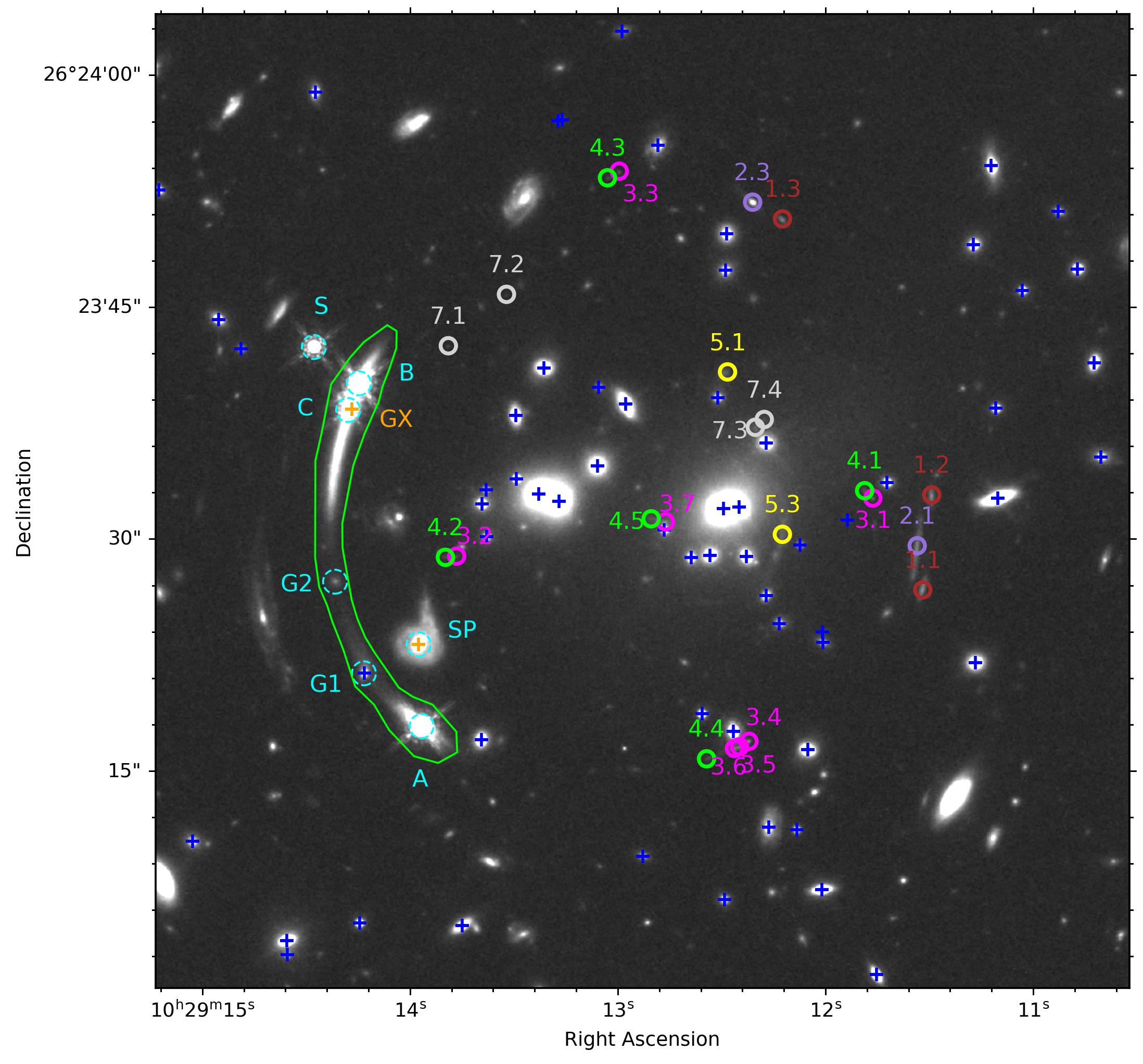} 
\caption{HST/WFC3 ($\sim$1 arcmin$^2$) image of SDSS1029 in the F160W band. The 26 spectroscopically-confirmed multiple images included in our analysis are marked with circles, color-coded per system. 
Spectroscopic cluster members are marked with blue crosses. The two line-of-sight galaxies included in our multi-plane strong-lensing modeling are highlighted with orange crosses.
The seven objects for which the light distribution is modeled are indicated with cyan dashed circles, and labeled following Section \ref{sec:LLM}.
The \textit{arc mask}, which encompasses the light of the three QSO multiple images and the host galaxy, is shown in green.
} \label{sdss1029_fig}
\end{figure*}

The lens cluster SDSS~J1029$+$2623 (hereafter SDSS1029) is of particular interest due to its potential application as a cosmological probe \citep{Moresco2022} and for studies of the strongly-lensed QSO and its host galaxy, at a relatively high-redshift \citep{Ding2017b}. At a redshift of $z=0.588$, this is one of the few presently known lens galaxy clusters producing multiple images (three, labeled as A, B, and C) of a background ($z = 2.1992$) QSO \citep[][]{Inada2006, Oguri2008}, with a measured time delay between the multiple images A and B of $(744~\pm~10)$ days \citep[][]{Fohlmeister2013}. 
In addition, the QSO host galaxy is lensed into a $\sim22\arcsec.5$ long tangential arc (see Figure \ref{sdss1029_fig}), thus an ideal testbed for surface-brightness modeling studies at galaxy-cluster scales.
The first \emph{position-based} strong-lensing analysis of SDSS1029 was presented by \citet{Oguri2013}, using high-resolution, multi-band HST observations. 
More recently, an improved \emph{position-based} lens model was presented by \citet[][\citetalias{Acebron2022} hereafter]{Acebron2022} by exploiting new spectroscopic observations taken with the Multi Unit Spectroscopic Explorer \citep[MUSE,][]{Bacon2014}, mounted on the Very Large Telescope (VLT).

In this work, we present a new strong-lensing model of SDSS1029, including as observables both the positions of \emph{pointlike} multiple-image systems, that span a wide redshift range, and the extended surface-brightness distribution of the QSO host galaxy, for the first time in such a complex and large-scale system. 
This work is thus set to provide a foundation for the extended surface-brightness modeling of multiply-imaged sources in galaxy clusters, paving the way for a new generation of cluster strong-lensing models.

This paper is organized as follows. In Section \ref{sec:data}, we concisely describe the HST imaging and VLT/MUSE spectroscopic data of SDSS1029. Section \ref{sec:SLM} provides an overview of the strong lensing modeling method and software used. Our results are presented and discussed in Section \ref{sec:results}. Finally, we draw our conclusions in Section \ref{sec:conclu}. 
Throughout the paper, we assume a flat ${\Lambda\mathrm{CDM}}$ cosmology with $H_0 = 70$ $ \mathrm{km~s^{-1}~Mpc^{-1}}$, and matter density $\mathrm{\Omega_{m}}=0.3$. In this cosmology, $1\arcsec$ corresponds to a physical scale of 6.62 kpc at the cluster redshift ($z=0.588$) and of 8.27 kpc at the redshift of the strongly-lensed QSO ($z=2.1992$).
Magnitudes are given in the AB system \citep{Oke1974}. Statistical uncertainties are quoted as the 68\% confidence level, unless otherwise noted.

\begin{table*}
  \renewcommand\arraystretch{1.2}

  \setlength\tabcolsep{0.4em}
  \centering
  \caption{Best-fit (BF) and marginalized (M) parameter values with the associated 1-$\sigma$ uncertainties for the light models of the seven perturber objects in SDSS1029, performed in the HST/F160W band. The $x$ and $y$ coordinates are given in arcseconds with respect to the position of the Brightest Cluster Galaxy, at R.A. = 157.3020471 degrees, Decl. = 26.3922094 degrees. The position angle, $\theta_{\rm PA}$, is measured counter-clockwise from the $x$-axis. Values in square brackets are kept fixed. We note that, for G2, the axis ratios of each isothermal profile are allowed to vary independently.
}
  \begin{tabular}{ccccccccc}
    \hline
    \hline
    &\multicolumn{2}{c}{PSF (S)} &
    \multicolumn{2}{c}{PSF (A)} &
    \multicolumn{2}{c}{PSF (B)} &
    \multicolumn{2}{c}{PSF (C)} \\
    \cmidrule(r){2-3}\cmidrule(l){4-5}\cmidrule(l){6-7}\cmidrule(l){8-9}
    Parameter & BF value & M value & BF value & M value & BF value & M value & BF value & M value\\
    \hline
    $x$ [$\arcsec$] & $-26.508$ & $-26.508^{+0.001}_{-0.001}$ &  $-19.567$ & $-19.567^{+0.001}_{-0.001}$ & $-23.611$ & $-23.611^{+0.001}_{-0.001}$ & $-24.308$ & $-24.308^{+0.001}_{-0.001}$\\
    $y$ [$\arcsec$] & $10.421$ & $10.422^{+0.001}_{-0.001}$ & $-14.107$ & $-14.107^{+0.001}_{-0.001}$ & $8.043$ &  $8.043^{+0.001}_{-0.001}$ & $6.355$ & $6.355^{+0.001}_{-0.001}$ \\
    $A$  & $582.6$ & $582.4^{+1.0}_{-1.0}$ & $1347.7$ & $1347.8^{+1.2}_{-1.2}$ & $1533.4$ & $1533.7^{+1.2}_{-1.2}$ & $495.1$ & $495.1^{+1.1}_{-1.0}$\\
    \hline
  \end{tabular}
  
\vspace{0.4cm}

    \begin{tabular}{ccccccccc}
    \hline
    \hline
    &\multicolumn{2}{c}{Chameleon (G1)} &
    \multicolumn{2}{c}{Chameleon (G2)} &
    \multicolumn{2}{c}{Chameleon (SP)} &
    \multicolumn{2}{c}{S\'ersic profile (SP)} \\
    \cmidrule(r){2-3}\cmidrule(l){4-5}\cmidrule(l){6-7}\cmidrule(l){8-9}
    Parameter & BF value & M value & BF value & M value & BF value & M value & BF value & M value\\
    \hline
    $x$ [$\arcsec$] & $[-23.274]$ & $[-23.274]$ & [$-25.166$] & [$-25.166$] & $-19.760$& $-19.760^{+0.002}_{-0.002}$ & $-19.760$& $-19.760^{+0.002}_{-0.002}$\\
    $y$ [$\arcsec$] & $[-10.671]$ & $[-10.671]$ & [$-4.735$] & [$-4.735$] & $-8.803$ & $-8.803^{+0.002}_{-0.002}$ & $-8.803$ & $-8.803^{+0.002}_{-0.002}$\\
    $q_{\rm L}$  & $0.49$ & $0.56^{+0.15}_{-0.13}$ & $0.77/0.44$ & $0.60^{+0.13}_{-0.16}/0.56^{+0.15}_{-0.16}$ & $0.61$ & $0.65^{+0.04}_{-0.02}$ & $0.82$& $0.82^{+0.03}_{-0.07}$\\
    $\theta_{\rm PA}$ [rad] & $1.44$ & $1.41^{+0.15}_{-0.16}$ & $2.40$ & $1.50^{+0.66}_{-0.70}$ & $1.31$& $1.33^{+0.03}_{-0.02}$ & $2.80$ & $2.79^{+0.02}_{-0.02}$ \\
    $I_0$  [mag arcsec$^{-2}$] & $0.030$ & $0.034^{+0.019}_{-0.006}$ & $0.006$ & $0.034^{+0.130}_{-0.022}$ & $0.286$ & $0.293^{+0.03}_{-0.02}$ & -- & -- \\
    $w_{\rm c}$ [$\arcsec$] & $0.012$ & $0.012^{+0.032}_{-0.009}$ & $0.065$ & $0.34^{+0.31}_{-0.21}$ & $0.001$ & $0.006^{+0.007}_{-0.004}$ & -- & -- \\
    $w_{\rm t}$ [$\arcsec$] & $0.547$ & $0.446^{+0.160}_{-0.170}$ & $0.916$ & $0.66^{+0.24}_{-0.30}$ & $0.159$ & $0.155^{+0.015}_{-0.013}$ & -- & -- \\
    $I_{\rm eff}$ [mag arcsec$^{-2}$] & -- & -- & --  & --  & -- & -- & $0.013$ & $0.013^{+0.001}_{-0.001}$\\
    $R_{\rm eff}$ [$\arcsec$] & -- & -- & --  & --  & -- & -- & $1.713$ & $1.730^{+0.060}_{-0.050}$ \\
    $n$ & -- & -- & --  & --  & -- & -- &  $0.98$ & $0.99^{+0.06}_{-0.06}$ \\

    \hline
  \end{tabular}
  
  \label{tab:light_models}
\end{table*}

\section{Data} \label{sec:data}
This section briefly summarizes the photometric and spectroscopic data sets used in this work. We refer the reader to \citetalias{Acebron2022} for a detailed overview.

\subsection{HST Imaging} \label{sec:HST}
We use archival HST high-resolution, multi-band imaging (GO-12195; P.I.: Oguri), from the \textit{Advanced Camera for Surveys} (ACS, 2 and 3 orbits in the ACS/F435W and ACS/F814W bands, respectively) and the \textit{Wide Field Camera 3} (WFC3, 2 orbits in the WFC3/F160W band). The observations and data reduction process are described in \cite{Oguri2013}. In this work, we mainly make use of the WFC3/F160W band\footnote{The F160W science and weight frames are downloaded from the \href{https://hla.stsci.edu}{HLA} and geotranned to the World Coordinate System (WCS) adopted in \citetalias{Acebron2022}.}, which provides the highest signal-to-noise ratio imaging of the strongly-lensed QSO host galaxy (see Figure \ref{sdss1029_fig}). With a pixel scale of $0\arcsec.03$, we are fully exploiting the power of these high-resolution HST observations (and showing the promise to carry out similar analyses with JWST data).
The error map associated with the HST F160W image is computed from the weight and the science frames (in units of counts per second) and includes the contribution from both the background and the Poisson noise. The total error value for each pixel, $i$, is thus:
\begin{equation}
  \sigma_{{\rm tot,}i} = \sqrt{1/w_{i} + d_{i}/t \times G},
\label{eq:errormap}
\end{equation}
where $w_{i}$ and $d_{i}$ denote the intensity of the pixel $i$ in the weight and science frames, respectively, $t$ is the exposure time and $G$ is the charge-coupled device (CCD) gain.

First, the positions and magnitude values of all the objects detected within the F160W field of view are measured with the software \texttt{Source Extractor} \citep{Bertin1996}. The luminosities of the member galaxies are measured using the HST F160W Kron magnitudes since they are a good proxy of their stellar mass \citep[see Figure 7 in][]{Grillo2015}. In a second step, the positions and magnitude values of 19 cluster members and a foreground spiral galaxy, that are affected by the presence of bright, angularly close neighbors, are refined with the \texttt{GALFIT} software \citep[][]{Peng2010}. The F160W magnitude values are given in Appendix~C in \citetalias{Acebron2022}. 

One key ingredient in our extended strong-lensing modeling is the point spread function (PSF). 
We follow the same PSF construction procedure as in \citet{Ertl2023}. Thus, we refer the reader to that publication for a detailed description, providing a brief summary hereafter. 
The PSF is created from the F160W science image (with a pixel scale of $0\arcsec.03$) by combining several non-saturated stars in the field. The star cutouts are then centered at the brightest pixel, stacked, and normalized. In this work, we use two different PSFs.
The first PSF, including the diffraction spikes, is directly used for the light profile modeling of the quasar multiple images or other point sources, such as stars.
The second PSF is used instead for the convolution of the light modeling of the extended sources. Since the computational time significantly increases with an increasing PSF size, it is cropped at the first diffraction minimum of the Airy disc, resulting in a PSF cutout $\sim 35\%$ smaller than that of the first, full PSF.

\subsection{VLT/MUSE Spectroscopy} \label{sec:MUSE}
We also use high-quality spectroscopic observations of SDSS1029, recently obtained with the integral field spectrograph VLT/MUSE under the program 0102.A-0642(A) (P.I.: Grillo). The observations, data reduction process, and redshift measurement methodology are detailed in \citetalias{Acebron2022}. 
Briefly, the observations consist of a single pointing ($\sim$~1~arcmin$^2$) with a cumulative exposure time of $4.8$ hours on target. 
The full spectroscopic catalogue includes 127 reliable redshift measurements. Among these objects, one foreground galaxy ($z =0.511$), 57 cluster members ($0.57\le z \le 0.61$), one background galaxy ($z =0.674$), and 26 multiple images ($1.02\le z \le 5.06$) are included in the strong-lensing modeling, presented in the following Section. The coordinates, redshift, and F160W magnitude values of these galaxies are given in Tables 2 and C1 in \citetalias{Acebron2022}.

\section{Strong-Lensing Modeling} \label{sec:SLM}

We perform the strong-lensing modeling of the galaxy cluster SDSS1029 with the Gravitational Lens Efficient Explorer software \citep[\GLEE,][]{Suyu2010, Suyu2012}.
Following \citetalias{Acebron2022}, we include in the lens model a foreground spiral and a background galaxy that are identified as SP and GX, respectively, and with orange crosses in Figure \ref{sdss1029_fig}. 
While, in a first approximation, the total mass contributions of SP and GX were modeled at the redshift of the galaxy cluster in \citetalias{Acebron2022}, in this work, we go further and make use of the full multi-plane lensing formalism \citep{Blandford1986, Schneider1992, Chirivi2018}.
The ray-tracing equation is thus modified to account for the multiple deflections produced by $n$ lenses at different redshifts along the line of sight:

\begin{equation}
  \boldsymbol \beta = \boldsymbol \theta_{n}(\boldsymbol \theta_1) = \boldsymbol \theta_{1} - \sum_{i=1}^{n-1} \frac{D_{in}}{D_{n}} \boldsymbol{\hat{\alpha}} (\boldsymbol \theta_i, \boldsymbol \xi),
  \label{lens_eq_mp}
\end{equation}

\noindent where $\boldsymbol{\theta}_{n}$ and $\boldsymbol{\theta}_{1}$ represent the positions of the light ray on the $n$-th and first lens plane, respectively, while $\boldsymbol{\theta}_{i}$ is the position of the light ray on the $i$-th lens plane. $D_{in}$ and $D_{n}$ are the angular-diameter distances between the $i$-th and the $n$-th lens plane, and between the observer and the $n$-th plane, respectively. $\boldsymbol{\hat{\alpha}}(\boldsymbol \theta_i, \boldsymbol \xi)$ is the deflection angle that a light ray undergoes on the $i$-th plane, given a total mass distribution described by a set of model parameters, $\boldsymbol \xi$. 

Within \GLEE, the different mass components and the surface brightness of (perturbing) objects are modeled in a parametric way, i.e., by adopting parametrized total mass and light profiles. The surface-brightness distribution of multiply-lensed sources is instead modeled on a pixel grid.
The best-fit and marginalized values are determined through simulated annealing and Markov chain Monte Carlo (MCMC) techniques \citep{Dunkley2005, Foreman-Mackey2013}.

This work aims to reconstruct the lens total mass distribution by including as observables both the positions of the \emph{pointlike} multiple images, in Section \ref{sec:PLMod}, and the surface-brightness distribution of the strongly-lensed QSO host galaxy, in Section \ref{sec:ExtMod}.

\subsection{Total mass model} \label{sec:MassModeling}
Following the reference strong lensing model, labeled as \emph{Model~1} in \citetalias{Acebron2022}, and constructed with the software Lenstool \citep{Jullo2007}, the total mass distribution of SDSS1029 is modeled with two large-scale dark-matter halos, modeled as non-truncated dual pseudo-isothermal elliptical (dPIE) mass density profiles \citep{Eliasdottir2007, Suyu2010, Bergamini2019}. 
The model parameters associated with the dPIE mass density profile are the centroid coordinates, expressed in arcseconds with respect to a given reference position, the axis ratio and position angle, the Einstein radius, the core and truncation radius (where the value of the truncation radius is set to an arbitrarily high value for our case of non-truncated halos).
The sub-halo mass component is modeled with a highly-pure sample of 83 cluster members, $\sim70\%$ of which are spectroscopically confirmed in the inner $\rm 1~arcmin^2$ region of the cluster thanks to the VLT/MUSE observations. In the cluster member catalog, 20 cluster galaxies, all lying outside of the VLT/MUSE field of view, and six within the VLT/MUSE footprint but with an \emph{insecure} spectroscopic quality flag, are selected with the red-sequence method.
Cluster galaxies are modeled with circular dPIE mass distributions, with a vanishing core radius, and all but one are scaled with total mass-to-light ratios increasing with their HST F160W luminosities, consistent with the Fundamental Plane \citep{Faber1987, Bender1992}.  
Instead, the total mass parameters of the cluster galaxy with ID 1645 in Table C1 in \citetalias{Acebron2022} (angularly very close to several multiple images from systems 3 and 4) are free to vary.
As previously mentioned, the two galaxies along the line-of-sight, the foreground SP ($z=0.511$) and the background hidden galaxy GX ($z=0.674$), are considered at their corresponding redshifts in our multi-plane modeling.
The multi-plane methodology is especially relevant in the lensing analysis of SDSS1029, as GX and SP lie angularly very close to the strongly-lensed QSO, and thus limits possible biases arising from assuming that these objects are at the lens cluster redshift. This can be particularly important for the background GX, as single-plane models do not take into account the fact that the background objects are lensed themselves (resulting in a magnification and deflection of their positions). SP and GX are modeled with circular dPIE mass density profiles, and their mass parameters (Einstein radius and truncation radius) are free to vary.
Finally, we introduce an external shear component and the redshift of system 1 is allowed to vary during the optimization.
The total number of free parameters associated with the total mass parametrization of SDSS1029 is $N^{\rm mass}_{\rm params}=22$. 

\subsection{Position-based lens modeling} \label{sec:PLMod}
In this work, we use the sample of \emph{pointlike} multiple images presented in \citetalias{Acebron2022}, which consists of 26 multiple images spanning a wide redshift range from $1.02$ to $5.06$. All multiple-image systems are spectroscopically confirmed except for system 1 (in dark red in Figure \ref{sdss1029_fig}, and with a quality flag `\emph{insecure}'), thus its redshift value is optimized in the lens model.
Their positions are shown in Figure \ref{sdss1029_fig} and we refer the reader to Table 1 in \citetalias{Acebron2022} for further details.

As performed in typical \emph{position-based} strong lensing models of galaxy clusters \citep[see e.g.][]{Jullo2007, Grillo2015, Caminha2017, Cerny2018, Acebron2019, Sharon2020, Diego2020, Bergamini2021, Granata2022, Jauzac2021}, the best-fitting values of the model parameters are found by minimizing on the image plane the distance between the observed, $\boldsymbol{\theta}^{\rm obs}$, and model-predicted, $\boldsymbol{\theta}^{\rm pred}$, positions of the multiple images, given a set of model parameters, $\boldsymbol{\xi}$ via the \emph{pointlike} chi-square function, $\chi^2_{\rm PL}$, defined as:

\begin{equation}
\chi^2_{\rm PL}=\sum_{j=1}^{N_{\rm fam}}\sum_{i=1}^{N_{\rm img}^{j}}\left(\frac{\left|\boldsymbol{\theta}^{\rm obs}_{ij}-{\boldsymbol{\theta}}^{\rm pred}_{ij}(\boldsymbol{\xi})\right|}{\sigma_{ij}}\right)^2,
\label{Eq:chi2PL}
\end{equation} 
where $N_{\rm img}$ and $N_{\rm fam}$ represent the number of \emph{pointlike} multiple images and families, respectively. $\sigma_{ij}$ is the positional uncertainty of the observed images. Following \citetalias{Acebron2022}, we adopt a positional uncertainty of $0.\arcsec25$ for most images, but we increase it to $0.\arcsec50$ for the multiple image systems 2 and 5, as their diffuse morphologies in the HST imaging entail a less reliable estimate of their positions. 

\begin{figure}
\centering
\includegraphics[width=\columnwidth]{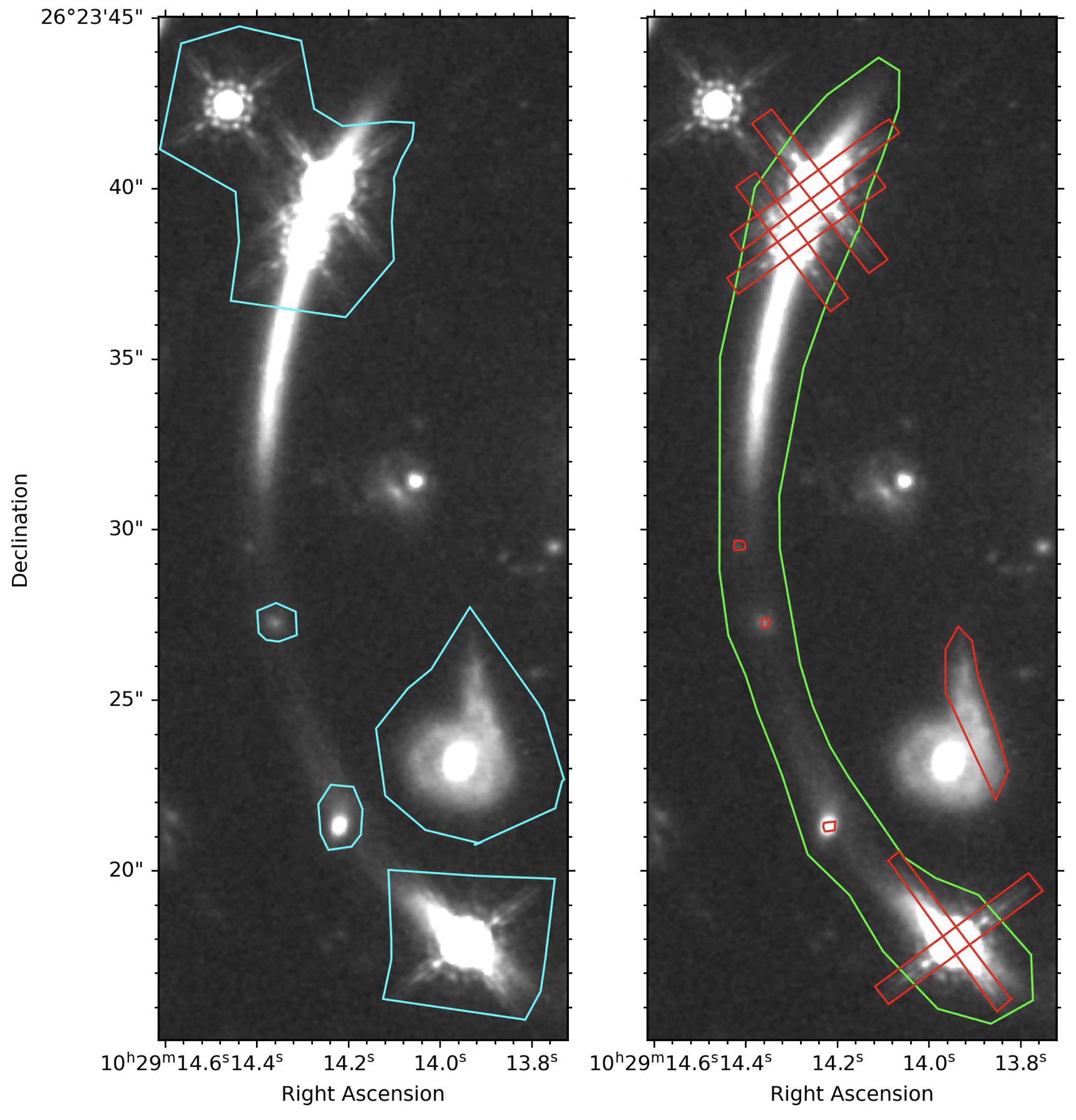} 
\caption{Zoom-in around the strongly-lensed QSO host galaxy in the WFC3/F160W band. \textbf{Left:} Cyan regions show the \textit{lensmasks}, encompassing the seven luminous objects for which the light distribution is modeled. In total, the \textit{lensmasks} contain $\sim 9.3 \times 10^4$ HST pixels. \textbf{Right:} Red regions indicate the areas where the observational uncertainties associated with the science image are boosted (by a factor of 10) when modeling the extended surface brightness of the lensed arc (see Section \ref{sec:ExtMod}). The number of boosted pixels represents $\sim 20\%$ of the \emph{arc mask} (in green).
} \label{ErrMasks}
\end{figure}

\begin{figure*}
\centering
\includegraphics[width=0.85\linewidth,trim=140mm 0mm 100mm 0mm, clip=true]{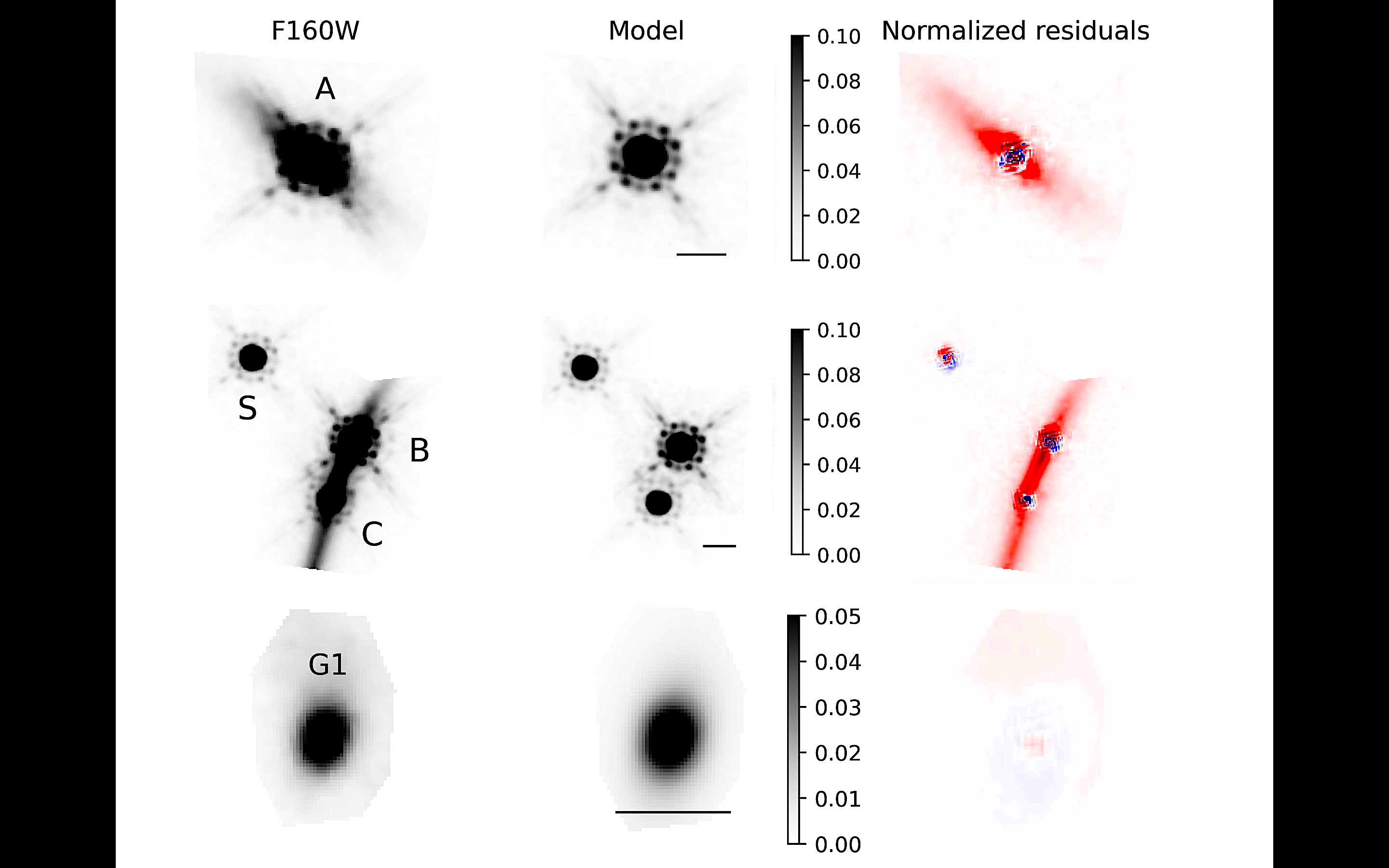}\\
\includegraphics[width=0.85\linewidth,trim=140mm 120mm 100mm 0mm, clip=true]{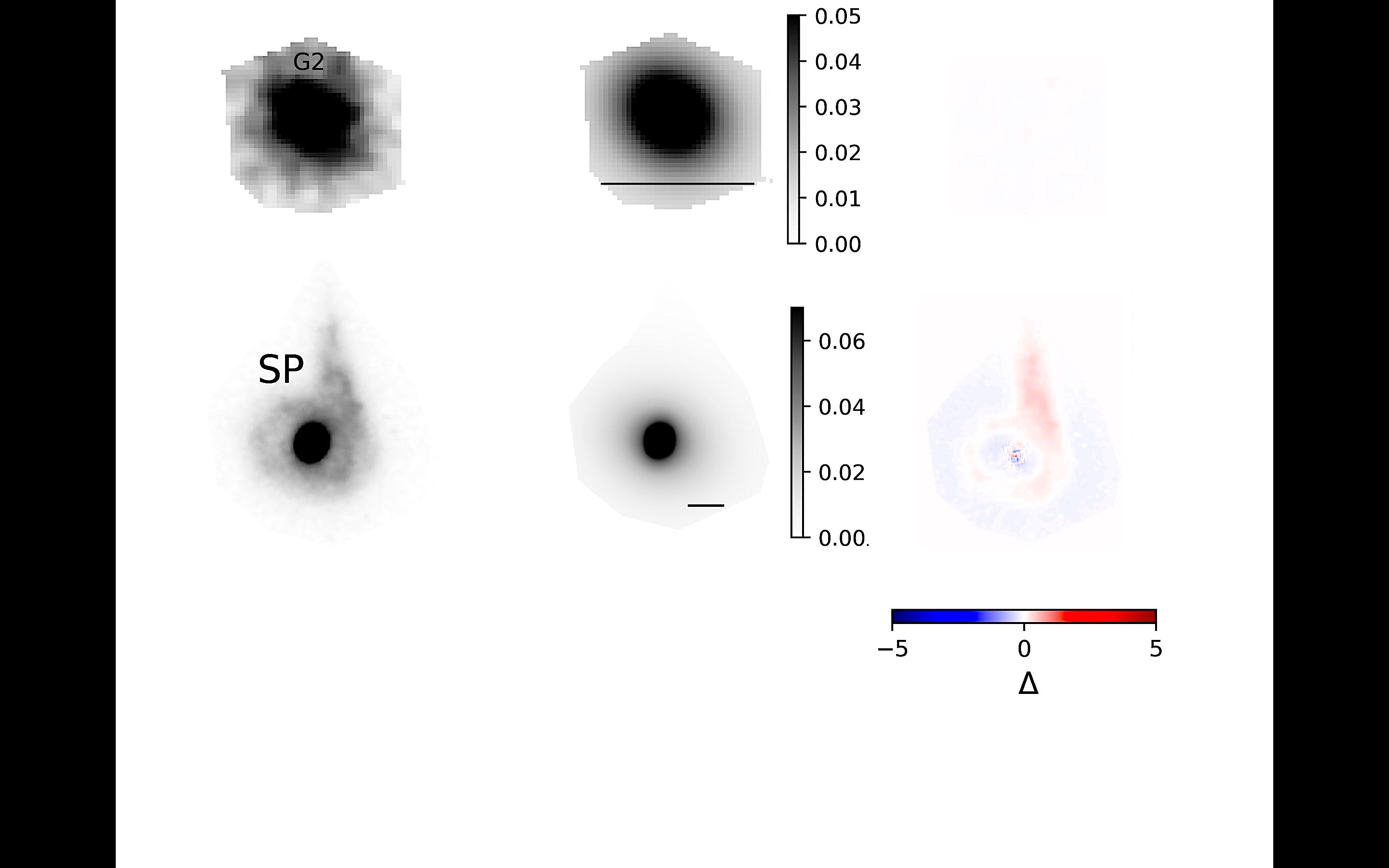}
\caption{
 Best-fit models for the surface-brightness distribution of the seven perturbing objects (following the labeling from Figure \ref{sdss1029_fig}). From left to right: the observed HST F160W band imaging, the best-fit model of their surface-brightness distribution, and the normalized residuals in a range between $-5\sigma$ to $5\sigma$. Angular scales of $1\arcsec$ are shown in the middle panels. North is up and East is left.
} \label{lightmodels}
\end{figure*}

\subsection{Lens Light Modeling} \label{sec:LLM}
Given that several objects are angularly very close to the lensed QSO host galaxy (see Figure \ref{sdss1029_fig}), it is necessary, as a first step, to subtract their light contamination before conducting the extended surface-brightness modeling in Section \ref{sec:ExtMod}. Specifically, we model the surface-brightness distribution of seven perturber objects, which are identified in cyan in Figure \ref{sdss1029_fig}. These are: one star (S), the three QSO multiple images (A, B, and C), and three galaxies (G1, G2, and SP). While the mass contribution of the hidden galaxy GX is included in our multi-plane lens model, we do not model its surface-brightness distribution. As shown in \citet[][see their Figure 4]{Acebron2022}, GX is a small and faint galaxy, showing [\ion{O}{ii}] and [\ion{O}{iii}] emission lines, reflecting the blue and star-forming nature of this object. While GX is scarcely visible in the F435W and F814W HST filters, it is not in the reddest band, F160W.

The surface-brightness distribution of these objects is modeled in the F160W band using one or multiple parametric light profiles within the \emph{lensmasks}. The latter are illustrated as the cyan regions in the left panel of Figure \ref{ErrMasks}, and contain in total $\sim 9.3 \times 10^4$ pixels. The adopted parametric light profiles for the different objects are described hereafter.\\

- \textit{The star and quasar multiple images}\\
The star and each QSO multiple image are directly modeled as a point source, represented by the PSF described in Section~\ref{sec:HST}.
There are three free parameters associated with the PSF light component, as we allow the central positions ($x$, $y$) and flux amplitudes ($A$) to vary.\\

- \textit{The galaxies G1 and G2}\\
For the early-type galaxies G1 and G2, we adopt the Chameleon light profile, which is defined as the difference between two non-singular isothermal elliptical profiles, with different core radii, but the same normalization. 
The Chameleon profile matches well the S\'ersic profile \citep{Sersic1963}. In particular, the difference is only of a few percent for radii between $0.5$ and $3.0$ effective radii \citep{Dutton2011, Suyu2014}. It is defined, in Cartesian coordinates ($x$, $y$), as:

\begin{equation} \label{eq:chameleon}
\begin{split}
        I_{\rm C} (x, y) &\equiv \frac{I_0}{1 + q_{\rm L}} \left[ \frac{1}{\sqrt{x^2+y^2/q_{\rm L}^2 + 4w_{\rm c}^2/(1+q_{\rm L}^2)}} \right. \\
         & \quad \quad \quad \quad \left. - \frac{1}{\sqrt{x^2+y^2/q_{\rm L}^2 + 4w_{\rm t}^2/(1+q_{\rm L}^2)}} \right],
\end{split}
\end{equation}

\noindent where $I_0$ is the flux amplitude normalization, $q_{\rm L}$ represents the axis ratio, and $w_{\rm c}$ and $w_{\rm t}$ are the core radii, with $w_{\rm t}>w_{\rm c}$. 

To model the light distribution of each of G1 and G2, we use a Chameleon profile with common, fixed centroids and position angles for the two isothermal profiles of the Chameleon.
In the case of G1, the axis ratios of the two isothermal profiles are linked, while for G2, the values of the axis ratio of each profile are free to vary independently (thus adding further flexibility in the modeling). \\ 

- \textit{The spiral galaxy SP}\\
Finally, we also model the light distribution of the foreground, bright spiral galaxy SP, at a projected distance of $\sim 4 \arcsec$ from the lensed arc. Being the most distant light perturber, and given its morphological complexity, the aim is to minimize its light contamination rather than perfectly modeling its surface-brightness distribution. 

The light distribution of SP is modeled with a combination of two different light profiles. We adopt a Chameleon light profile to model the bulge component, where the two isothermal profiles have common centroids, axis ratios and position angles. 

The disk component of the spiral is instead modeled with a S\'ersic profile \citep{Sersic1963}, that is defined as:

\begin{equation}
        I_{{\rm S}} (x, y) \equiv I_{\rm eff} \exp \left[-b_n \left\{\left(\frac{\sqrt{x^2 + y^2/q_{\rm L}^2}}{R_{\rm eff}} \right)^{1/n} - 1 \right\} \right],
\label{Eq:Sersic}
\end{equation}

\noindent where $I_{\rm eff}$ is the amplitude at the effective radius $R_{\rm eff}$, and $n$ is the S\'ersic index \citep{Sersic1968}. The normalizing factor, $b_n$, depends on $n$, so that $R_{\rm eff}$ corresponds to the half-light radius \citep{Ciotti1999}. The centroid of the S\'ersic profile is linked to that of the bulge light component, while the remaining parameters are left free to vary independently.\\

The best-fit parameters values for each light model are first estimated through a simulated annealing technique that minimizes the 
difference in each pixel between the observed intensity, $I^{\rm obs}$ and the predicted one, $I^{\rm pred}$, convolved with the PSF. 
The PSF used for the quasar multiple images and the star has a pixel scale of $0.03\arcsec$ and covers $\sim 4.2\arcsec \times 4.2\arcsec$, so as to include the diffraction spikes. The PSF used for the light modeling of the three galaxies is cropped to limit the increase in computational time (see Section \ref{sec:HST}). 
The $\chi^2_{\rm light}$ function is defined as: 
\begin{equation}
\chi^2_{\rm light}=\sum_{i=1}^{N_{\rm pix}}\left(\frac{I^{{\rm obs}}_i-I^{{\rm pred}}_i \otimes \rm{PSF}}{\sigma_{{\rm tot}, i}}\right)^2, 
\end{equation} 
where $\sigma_{{\rm tot}, i}$ represents the value of the total error map at the pixel $i$, as defined in Section \ref{sec:HST}.
The median values and the associated $1\sigma$ (68\% confidence level) uncertainties are instead extracted from MCMC chains with a total number of $2\times 10^6$ steps, with an acceptance rate value of $\sim25\%$, and rejecting the first 10\% burn-in steps. 

The best-fit and marginalized values with the associated $1\sigma$ uncertainties for the light distribution of the seven perturber objects are reported in Table \ref{tab:light_models}.
The best-fit surface-brightness models and the resulting normalized residuals are shown in Figure \ref{lightmodels}. 

\begin{figure*}
\centering
\includegraphics[width=\linewidth]{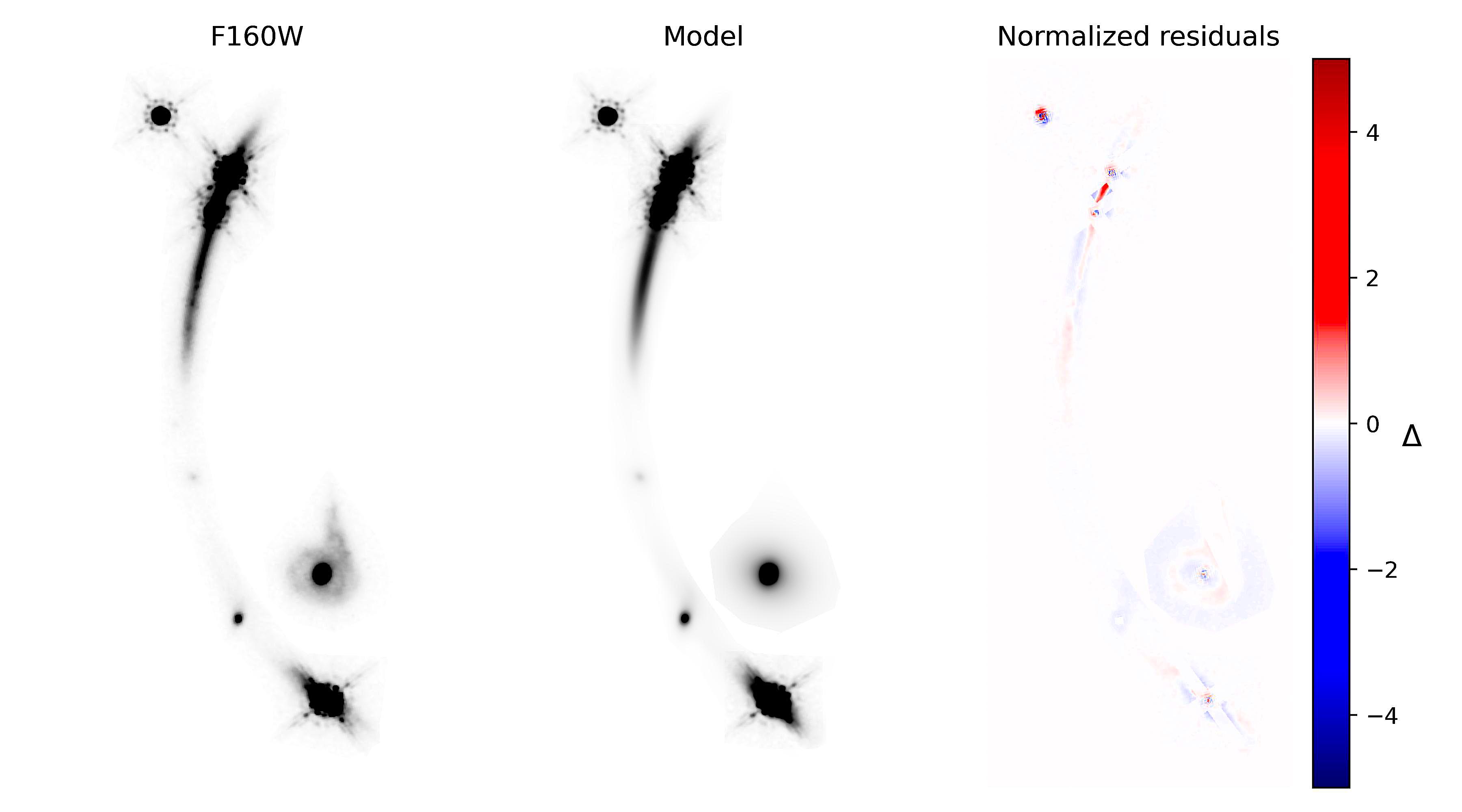} 
\caption{Best-fit surface-brightness reconstruction of the QSO host galaxy at $z=2.1992$. From left to right: the observed HST F160W band image over a $\sim 11\arcsec \times 29\arcsec$ field-of-view (or $ \sim 73 ~ \rm kpc \times 192 ~\rm kpc $ at the galaxy cluster redshift), the best-fit model, and the normalized residuals in a range between $-5\sigma$ to $5\sigma$. North is up and East is left.
} \label{arcmodel}
\end{figure*}

\subsection{Extended surface-brightness modeling} \label{sec:ExtMod}
In this Section, we describe the surface-brightness modeling of  multiply-lensed extended sources with \GLEE, that we apply in this work to the QSO host galaxy, strongly-lensed by SDSS1029 into a $\sim22\arcsec.5$ long tangential arc (see Figure \ref{sdss1029_fig}), with a diffuse morphology in the HST imaging and no clear substructures. 
We perform the extended lens modeling in the HST F160W band, where the signal-to-noise ratio of the QSO host galaxy is the highest and where the impact of any contamination from the galaxy GX is minimal (see the second row in Figure \ref{lightmodels}).  

The first step consists in creating the \emph{arc mask} which defines the region of pixels containing the observed QSO multiple images and the host galaxy. The resulting \emph{arc mask}, that is used in the extended surface-brightness modeling, is shown as the green region in Figures \ref{sdss1029_fig} and \ref{ErrMasks}, and encapsulates $N_{\rm pix}^{\rm arcmask}=77700$ pixels. This is the first time that an analysis of this kind is performed in such a complex and large-scale system, overcoming computational challenges. A single image-plane, extended surface-brightness optimization of SDSS1029 requires $\sim 100$ GB of RAM for $\sim 5$ days\footnote{Adopting 1000 steps at a given temperature, an initial and final temperature of 1 and 0.1, respectively, and a temperature cooling factor of 1.1.} on a single core (with 2195.905 CPU MHz), while one MCMC chain, with approximately $10^6$ total steps and an acceptance rate between $20\%$ and $30\%$, can require said RAM allocation for several months.
The computational time needed to perform analyses of this kind can be reduced with improved PSF convolution methods, the use of Graphics Processing Units (GPUs), or novel and highly efficient modeling techniques \citep[see e.g.,][]{Galan2022, Lombardi2024}.

The light intensity of each of the three QSO multiple images, obtained in Section \ref{sec:LLM}, is the superposition of the amplitude from the QSO and the contribution from its host galaxy. To properly model the light distribution of the host galaxy, we decrease the values of the QSOs amplitude parameter by $\sim 30 \%$ \citep[based on the well-tested modeling procedure detailed in][which has been applied extensively on similar QSO systems strongly-lensed by single galaxies]{Ertl2023}.
During the optimization process, the amplitudes of the QSO multiple images are allowed to vary, while the remaining parameters of the light profiles associated with the objects S, G1, G2, and SP, as well as the QSOs centroids, are fixed to their best-fit values previously obtained (see Section \ref{sec:LLM}).

Since the QSO multiple images and the galaxies G1 and G2 are much brighter than the light of the underlying arc, the residuals from the light models in the most central regions of the perturber objects can  overwhelm the signal from the arc \citep{Ertl2023}.
Therefore, before modeling the light distribution of the arc, the error map, $\sigma_{\rm tot}$, is boosted by a factor of $10$ in the regions with high residuals from the light distribution models (see Section \ref{sec:LLM}). 
We also boost the errors in the regions containing an extremely faint galaxy north of G2, the tail of SP (as it is unclear from the HST color image that it is associated with the foreground spiral galaxy), and the diffraction spikes of the QSO multiple images. 
Due to the rescaling of the QSO flux amplitudes during the extended surface-brightness modeling, larger uncertainties on the PSF diffraction spikes help limiting the impact of non-perfectly modeled areas on the reconstructed source.
The regions with boosted uncertainties, delimited in red in the right panel of Figure \ref{ErrMasks}, contain $N_{\rm boosted~pix}^{\rm arcmask}=15768$ pixels, i.e., $\sim 20\%$ of the \emph{arc mask} (in green). These regions are thus effectively not considered during the surface-brightness modeling.

The pixels within the \emph{arc mask} are mapped onto the source plane through the ray-tracing equation, where the surface-brightness of the source is reconstructed on a square grid of $40 \times 40$ pixels. 
A curvature form of regularization is applied on the source surface-brightness pixels, through the \emph{regularization constant}, $\lambda$ that represents the strength of the regularization \citep{Suyu2006}. 
The optimal value of $\lambda$ is obtained by solving Equation 20 in \citet{Suyu2006}, where solutions of $\lambda$ resulting in sources with an irregular surface-brightness distribution are penalized.
The pixel grid is then mapped back onto the image plane and the lens mass parameters are optimized by minimizing the difference between the predicted light intensity, $d^{\rm pred}$, and the observed one, $d^{\rm obs}$, in each pixel $i$ through the extended chi-square function, $\chi^2_{\rm EXT}$, that is written as:

\begin{equation}
\chi^2_{\rm EXT}=(\boldsymbol{d^{\rm obs}}-\boldsymbol{d^{\rm pred}})^T C_D^{-1}(\boldsymbol{d^{\rm obs}}-\boldsymbol{d^{\rm pred}}), 
\label{Eq:chi2ext}
\end{equation} 
where $\boldsymbol{d^{\rm obs}}$ and $\boldsymbol{d^{\rm pred}}$ are vectors with dimensions $N_{\rm pix}^{\rm arcmask}$ (the number of pixels in the \emph{arc mask}), and $C_D$ is the image covariance matrix \citep{Suyu2006}.

The final number of model constraints, $N_{\rm con}$, optimized model parameters, $N_{\rm free}$, and thus the number of degrees of freedom (dof), $N_{\rm dof}$, related to the extended strong-lensing model are computed as:

\begin{equation}
\begin{split}
\begin{gathered}
N_{\rm con}= 2 \times N_{\rm img} + N_{\rm pix}^{\rm arcmask},\\
N_{\rm free}= 2 \times N_{\rm fam} + N^{\rm source}_{\rm pix} + N^{\rm mass}_{\rm params}+ N_{z, \rm s1} + N^{\rm light}_{\rm params}, \\
N_{\rm dof}=N_{\rm con}-N_{\rm free}.
\end{gathered}
\end{split}
\label{Eq:dofext}
\end{equation} 
$N^{\rm source}_{\rm pix}$ represents the effective number of pixels on the source plane, constrained by the regularization constant $\lambda$, and is equal to 318 for the best-fitting model. 
$N_{z, \rm s1}$ accounts for the freely optimized redshift of the multiple image system 1. $N^{\rm light}_{\rm params}$ denotes the number of free parameters related to the light modeling. In the extended lens model of the arc, only the values of the amplitude of the three QSO multiple images are allowed to vary, while those of all the other parameters describing the surface-brightness distribution of the perturbers (see Table~\ref{tab:light_models}) are fixed to their best-fit values. However, given that these best-fit values were obtained through previous optimization processes, we add them to the total number of free parameters related to the lens light modeling, of $N^{\rm light}_{\rm params}=37$ (see Table~\ref{tab:light_models}).
Thus, our extended lens model has a total of 77360 dof, compared to 17 in the previous \emph{position-based} model (\citetalias{Acebron2022}). 

The observables in our extended lens model include the positions of the \emph{pointlike} multiple images, and the surface brightness of the pixels in the F160W band that are fitted simultaneously. Thus, during the model optimization, it is the combination of Equations \ref{Eq:chi2PL} and \ref{Eq:chi2ext} that is minimized:

\begin{equation}
\chi^2_{\rm tot}=\chi^2_{\rm PL}+\chi^2_{\rm EXT}.
\label{Eq:chi2tot}
\end{equation} 

The values of the model parameters (summarized in Equation \ref{Eq:dofext}) are firstly estimated through a simulated annealing technique. We then run a first MCMC chain that is used to estimate the covariance matrix of the model parameters and to extract the starting point for the second MCMC chain.
The best-fit model, as well as the median values and the $1\sigma$ uncertainties are extracted from this final MCMC chain with approximately $10^6$ total steps, with an acceptance rate of $\sim$ 30\%, and removing the first 10\% burn-in steps. 

\section{Results and Discussion} \label{sec:results}
In this Section, we present the results of a new lens model of SDSS1029, that exploits the extended surface-brightness information of the strongly-lensed QSO host galaxy. 
The most reliable strategy to include as an observable the smooth, surface-brightness distribution of a strongly-lensed source (i.e. without clear substructures), is to carry out a full extended lens modeling.
We discuss the potential of our new, extended lens model to study the intrinsic properties of the source in Section \ref{sec:Host}. We also compare, in Section \ref{sec:TMD}, the cluster total mass distribution, and the mass parameters values of two subhalos angularly-close to the giant arc, with those from the previous \emph{position-based} lens model presented in \citetalias{Acebron2022}.

The best-fit extended strong lensing model yields a root mean square separation between the model-predicted and observed  positions of the \emph{pointlike} multiple images, shown in Figure \ref{sdss1029_fig}, of $\rm rms= 0\arcsec.63$, significantly worsening the fit compared to the reference \emph{position-based} lens model (\textit{Model 1}) in \citetalias{Acebron2022}, with $\rm rms= 0\arcsec.15$.  
This is explained by the fact that the number of observables from the extended surface-brightness modeling of the QSO host galaxy is several orders of magnitude larger than the number of observables from the positions of the \emph{pointlike} multiple images (38, see \citetalias{Acebron2022}), entailing that the former term dominates the contribution to the total chi-square function, $\chi^2_{\rm tot}$. Here, we focus on the reconstruction of the surface-brightness distribution of the strongly-lensed QSO host galaxy, and delegate to a future work the study of alternative combinations of the different types of observables. We note, however, that the positions of the QSO \emph{pointlike} multiple images are well reproduced, with a precision of $\rm rms= 0\arcsec.13$.

The best-fit surface brightness reconstruction of the strongly-lensed QSO host galaxy is shown in Figure~\ref{arcmodel}. The comparison between the left and middle panels demonstrates that our extended lens model reproduces extremely well the observed intensity and morphology of the QSO host galaxy in the HST F160W band.
From the normalized residual map (right panel in Figure~\ref{arcmodel}), the imperfect light model near the centers of the pointlike perturbers (the star and the three QSO multiple images, see Figure~\ref{lightmodels}) can be noticed \citep[see also][]{Wong2017, Grillo2020}. These large residuals, in a relatively small area, do not affect our results since the errors in these regions have been boosted \citep[see Section \ref{sec:ExtMod}, Figure~\ref{ErrMasks}, and][]{Suyu2012b}.
We finally note from the normalized residual map that the observational uncertainties (see Section~\ref{sec:data}) are overestimated (where the contribution of the weight to the total error is of about $95\%$), translating into conservative errors on the values of the total mass model parameters.
We find that, when comparing our best-fit extended surface-brightness distribution model (see Figure~\ref{arcmodel}) with an `a posteriori' reconstruction from a \emph{position-based} GLEE lens model (see Figure~\ref{extreconstruction}), the normalized residuals are, as expected, a factor $\sim20$ better, on average, in the former case, where it is directly considered as an observable. This further supports the need to carry out extended lens models to robustly characterize the intrinsic properties of the strongly-lensed sources and measure the total mass of angularly-close subhalos. 

\begin{table}
    \centering
    \caption{Best-fit (BF) and marginalized (M) parameter values with the associated 1-$\sigma$ uncertainties for the S\'ersic light model of the reconstructed QSO host galaxy on the source plane. The adopted notation follows that of Table \ref{tab:light_models}.}
    \begin{tabular}{c|c|c}
         Parameter & BF value & M value \\
         \hline
         $x$ [$\arcsec$] & 1.640 & $1.641^{+0.001}_{-0.001}$ \\
         $y$ [$\arcsec$] & 1.314 & $1.315^{+0.001}_{-0.001}$ \\
         $q_{\rm L}$ & 0.82 & $0.82^{+0.01}_{-0.01}$ \\
         $\theta_{\rm PA}$ [rad] & 0.00 & $0.00^{+0.01}_{-0.01}$ \\
         $I_{\rm eff}$  [mag arcsec$^{-2}$] & 0.068 & $0.068^{+0.001}_{-0.001}$ \\
         $R_{\rm eff}$ [$\arcsec$] & 0.300 & $0.300^{+0.01}_{-0.01}$ \\
         $n$ & 0.60 & $0.60^{+0.01}_{-0.01}$ \\
         \hline
    \end{tabular}
    \label{tab:sersicfit}
\end{table}

\subsection{Physical properties of the quasar host galaxy} \label{sec:Host}
\begin{figure}
\centering
\includegraphics[width=0.98\columnwidth]{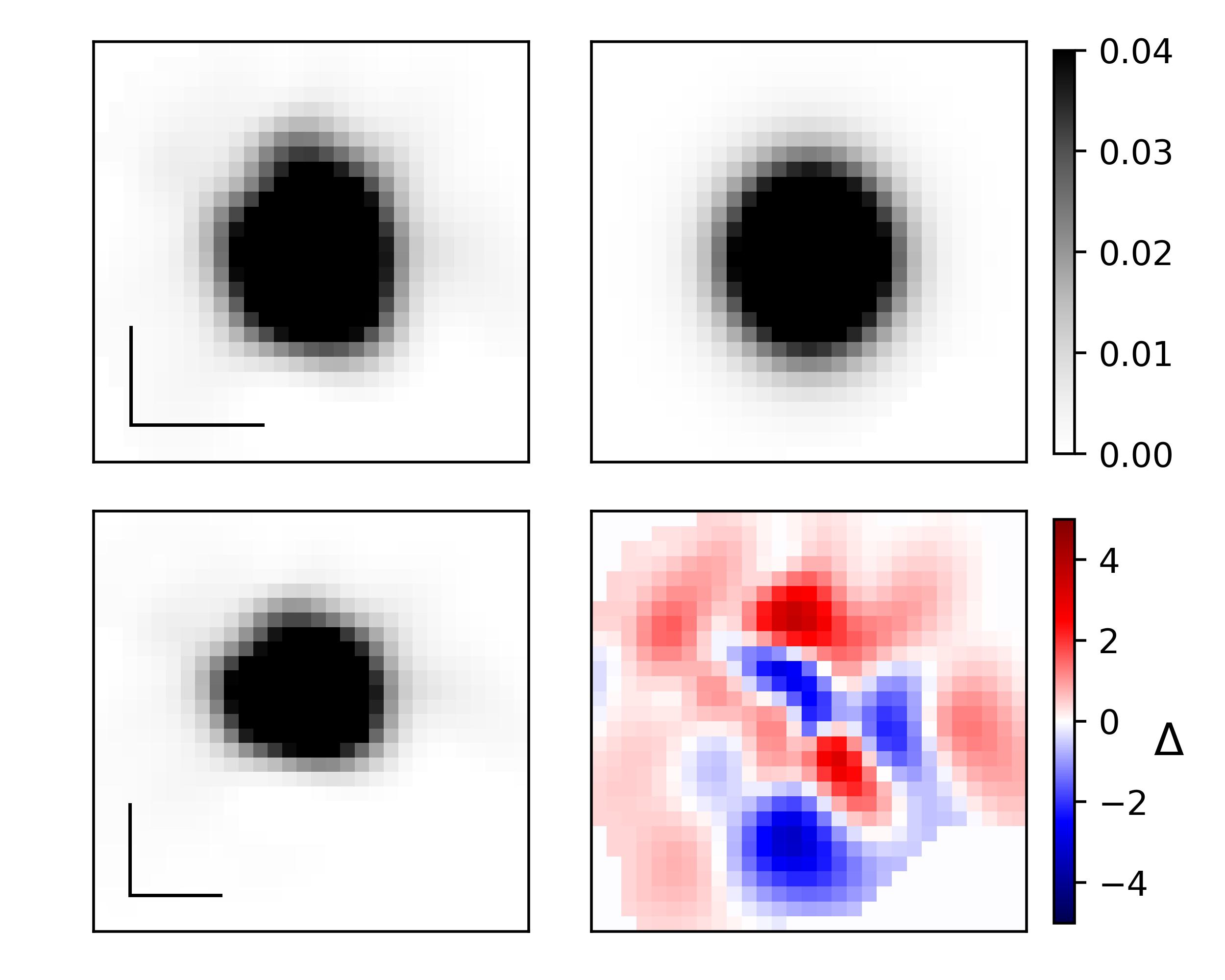} 
\caption{Best-fitting source surface-brightness reconstruction of the QSO host galaxy and modeling. \textbf{Top left:} Best-fitting source surface-brightness reconstruction of the QSO host galaxy on the source plane within an area of $\sim 3.5 ~ \text{arcsec}^{2}$, at $z=2.1992$. \textbf{Bottom left:} Reprojected source onto a rectangular, isotropic grid. The bars show the angular scales of $0.5\arcsec$ along the horizontal and vertical axes for both the reconstructed and reprojected source. \textbf{Top right:} Best-fitting surface-brightness distribution model of the reconstructed source (shown in the top left panel), obtained with a single S\'ersic profile. \textbf{Bottom right:} Normalized residuals in a range between $-5\sigma$ to $5\sigma$.
} \label{sourceQSO}
\end{figure}

The best-fitting source surface-brightness distribution is shown in the top left panel of Figure \ref{sourceQSO}, within an area probed on the source plane of $\sim 7 ~ \text{arcsec}^{2}$. The reconstructed source exhibits a single, compact, and smooth peak in its surface-brightness distribution.  
Within \GLEE, the source is reconstructed on a square grid, of $40 \times 40$ pixels in this case, where the pixel size along the $x$ and $y$-axes is different. We thus perform a flux-conserved projection of the reconstructed source plane surface-brightness map into a rectangular grid with the same pixel scale of $0\arcsec.08$ along the $x$ and $y$-axes. The reprojected source is shown in the bottom left panel of Figure \ref{sourceQSO}.
We estimate the statistical uncertainty from the lens modeling on the predicted intensity and morphology of the reconstructed source, estimated as the standard deviation from 100 random model realizations extracted from the final MCMC chain. We find that the relative statistical error on the predicted source intensity is small, with a mean value of $\lesssim 5\%$ over the extension of the source.

To infer the intrinsic properties of the background source, we fit its surface-brightness distribution with a S\'ersic profile with \GLEE, which adopts a Bayesian approach (see Section \ref{sec:LLM}) and takes into account the different pixel sizes along the $x$ and $y$-axes. We adopt flat priors on the model parameters and the error map is approximated as a constant, set to be the root mean square of the source intensity at the edges of the source, better reflecting the typical uncertainty map of the source intensity reconstruction \citep{Suyu2006}. The 1-$\sigma$ uncertainty values of the source intensity distribution are, in fact, quite uniform both inside and outside of the caustic \citep{Suyu2006}.
The best-fitting surface-brightness distribution model of the source is obtained with a single S\'ersic profile, and is shown in the top right panel of Figure \ref{sourceQSO}, while the normalized residuals are shown in the bottom right panel. The corresponding best-fitting and marginalized values of the model parameters are reported in Table \ref{tab:sersicfit}. 
We find that a combination of two S\'ersic profiles (with $n_1\sim1$ and $n_2\sim4$) yields a similar fit, and no significant improvement given the increased number of model parameters. 
From Equation \ref{Eq:Sersic}, we infer an intrinsic magnitude of the QSO host galaxy in the F160W band of $m_{\rm host} = 23.3 \pm 0.1$, where the quoted error reflects the statistical error from the lens model and the systematic uncertainty from the light model with a single or double S\'ersic parameterization (see Table \ref{tab:sersicfit}), showing the small dependence of the intrinsic magnitude estimate on modeling assumptions. 
We also measure, from the cumulative luminosity profile of the source, a half-light radius of the QSO host galaxy of (0.289 $\pm$ 0.003)$\arcsec$, which corresponds to a physical size of (2.39 $\pm$ 0.03) kpc at $z=2.1992$, where the error value includes the statistical uncertainty from the lens model and possible approximations from the reprojection of the source. 
Leveraging the large magnification values in lens cluster fields, this work demonstrates that cluster-lensed QSOs can provide a unique and complementary opportunity to extend the study of the relation between supermassive black-hole mass and host-galaxy properties to fainter magnitudes than those probed by non-lensed \citep[e.g.,][]{Peng2006, Treu2007, Bennert2011, Schramm2013, Park2015} and galaxy-scale-lensed \citep{Peng2006, Ding2017b} QSOs, at cosmological distances.

\subsection{Cluster Mass distribution} \label{sec:TMD}
\begin{figure}
\centering
\includegraphics[width=\columnwidth]{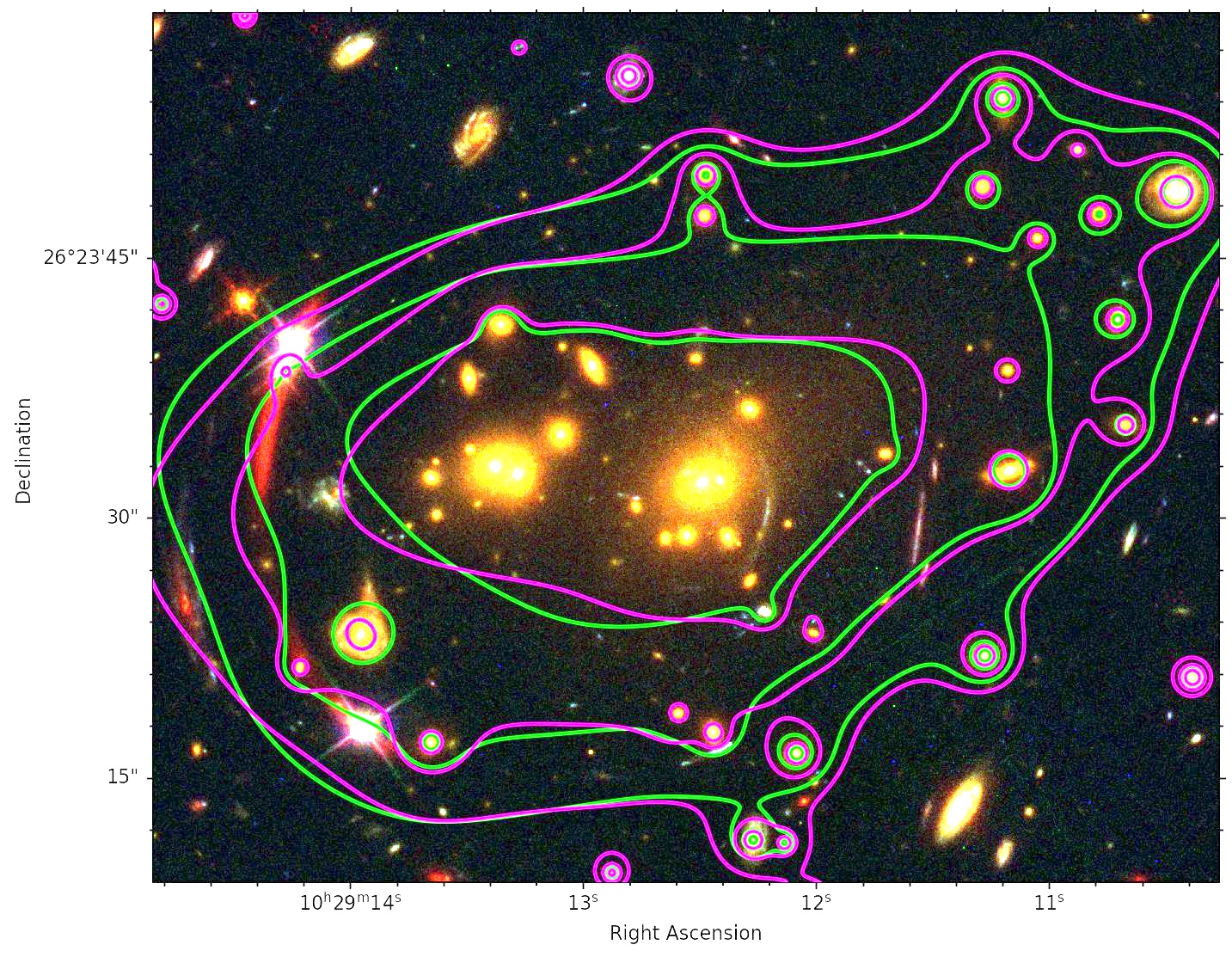} 
\caption{RGB image constructed with the HST passbands F435W (blue), F814W (green), and F160W (red) of the central region of SDSS1029, where we display the isodensity contours [0.65, 0.85, 1.35] of the convergence for $D_{\rm ls}/D_{\rm s}~=~1$. The green and magenta contours correspond, respectively, to the best-fit \textit{position-based} Lenstool model (\textit{Model~1} in \citetalias{Acebron2022}) and the \GLEE\ best-fit extended model presented in this work.
} \label{kappa}
\end{figure}

In Figure \ref{kappa}, we compare the reconstructed isodensity contours of the dimensionless projected surface mass density, or convergence, of SDSS1029 obtained from the \textit{position-based} Lenstool (considering the reference \textit{Model 1} from \citetalias{Acebron2022}) and the extended \GLEE\, lens models in green and magenta, respectively. We find a good agreement between the two total mass models, where the main differences (the values of the convergence differ between $\sim5\%$ to $\sim10\%$) arise around the newly-modeled extended QSO host galaxy in the \GLEE\, model or in the north-east region of the lens cluster core, where no multiple images have been identified (see also Figure \ref{sdss1029_fig}).

\begin{figure*}
\centering
\includegraphics[width=0.83\linewidth]{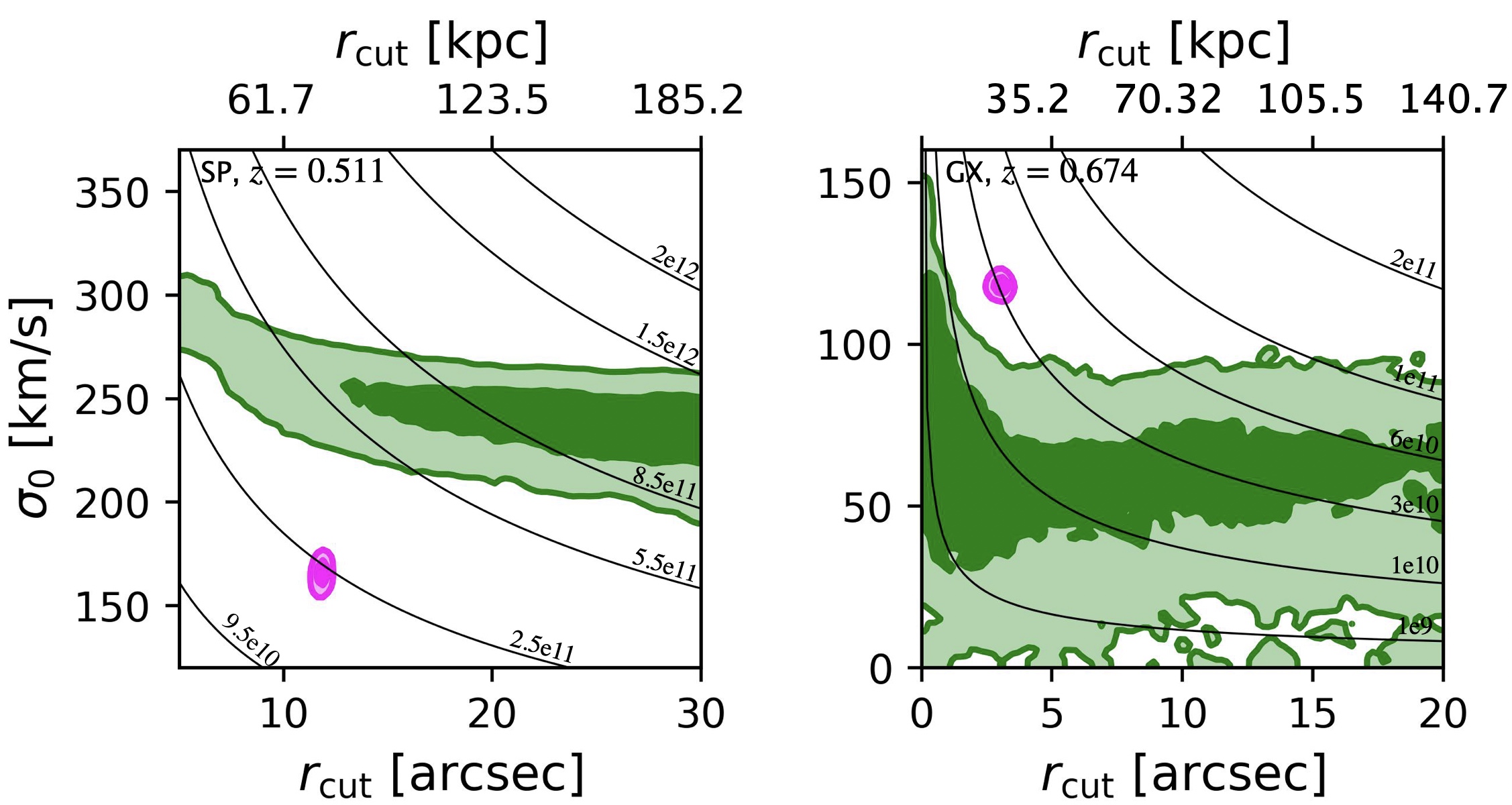} 
\caption{1- and 2-$\sigma$ confidence levels of the joint 2D probability distributions of the velocity dispersion and truncation radius of the foreground spiral galaxy (SP, \textbf{left}) and the hidden galaxy (GX, \textbf{right}) from the \textit{position-based} Lenstool and the extended \GLEE\, lens models in green and magenta, respectively. The black lines denote levels of equal total mass in solar mass units. 
In the adopted cosmology, $1\arcsec$ corresponds to a physical scale of 6.18 kpc and 7.04 kpc at the redshift of SP and GX, respectively. We note that both galaxies are modeled instead at the lens cluster redshift ($z=0.588$) in the \textit{position-based} Lenstool model by \citetalias{Acebron2022}.
} \label{SP_GX}
\end{figure*}

We also investigate the potential of our multi-plane, extended model to probe the total masses of the two line-of-sight galaxies, SP ($z=0.511$) and GX ($z=0.674$), that are individually modeled with circular dPIE mass density profiles. While GX sits on top of the lensed arc, SP lies at a projected distance of $\sim 4 \arcsec$ (see Figure \ref{sdss1029_fig}). Figure \ref{SP_GX} shows the comparison of the posterior distributions of the central velocity dispersion, $\sigma_0$, and the truncation radius, $r_{\rm cut}$, of the foreground spiral galaxy (SP, left panel) and the hidden galaxy (GX, right panel). The 1- and 2-$\sigma$ confidence levels from the \textit{position-based} Lenstool and the extended \GLEE\, lens models are shown in green and magenta, respectively. The reported values of $\sigma_0$ from our Lenstool model, corrected by the factor $\sigma_0 = \sqrt{3/2}~ \sigma_{\rm Lenstool}$, can be related to the Einstein radius of a singular isothermal sphere (SIS) mass density profile, for a lensed source at infinite redshift, from the relation
\begin{equation}
\theta_{\rm E, SIS} = \frac{4\pi \sigma_0^2}{c^2},
\label{Eq:thetaE_SIS}
\end{equation} 
where $c$ is the speed of light.
We can clearly remark that including the large number of observational constraints from the extended arc is crucial to reduce the degeneracies between the two model parameters as well as with the other mass components describing the complex mass structure of the main lens cluster \citep[see also][]{Wang2024}. From the \textit{position-based} Lenstool model, we estimate the median and 1-$\sigma$ uncertainties $\sigma_0^{\rm SP, ~LT} = 242^{+18}_{-18}$ km/s, $r_{\rm cut}^{\rm SP, ~LT}= 20.0 \arcsec^{+0.2}_{-14.3}$, and $\sigma_0^{\rm GX, ~LT} = 58^{+24}_{-27}$ km/s, $r_{\rm cut}^{\rm GX, ~LT}= 9.1\arcsec^{+12.8}_{-0.3}$ (\citetalias{Acebron2022}). Instead, the extended lens model with \GLEE\, yields $\sigma_0^{{\rm SP, ~}\GLEE} = 166^{+5.3}_{-5.5}$ km/s, $r_{\rm cut}^{{\rm SP, ~}\GLEE}= 11.8\arcsec^{+0.1}_{-0.1}$ and $\sigma_0^{{\rm GX, ~}\GLEE} = 118^{+0.2}_{-0.3}$ km/s, $r_{\rm cut}^{{\rm GX, ~}\GLEE}= 3.0\arcsec^{+0.2}_{-0.3}$.
These values are in strong disagreement, in particular for SP. This is highlighted in Figure \ref{SP_GX}, where we show different levels of equal total mass (black lines).
While we find a significantly different total mass value for SP, for GX the new parameter values, despite being located in a different region of the parameter space, provide total mass estimates that are consistent with the previous ones from \citetalias{Acebron2022}.
The differences between the results from the two analyses can be explained by the lack of observables in the \emph{position-based} Lenstool model in this cluster outer region, where only the three QSO \emph{pointlike} multiple image positions were considered (\citetalias{Acebron2022}).  

To confirm the robustness of our new extended lens model, it would be useful to compare the estimates of the total mass enclosed within a radius $r$ from our strong lensing analyses with that from an independent method.
We can carry out this test for SP, for which we can independently infer, by leveraging the high-quality integral-field spectroscopy from VLT/MUSE, its total dynamical mass from the gas kinematics.
Indeed, the total dynamical mass of SP can be derived via

\begin{equation}
M_{\rm dynamics}(r) = \frac{r~v_{\rm rot}^2(r)}{G},
\label{Eq:Mdyn}
\end{equation} 
where $G$ is the universal gravitational constant.
To measure the value of the asymptotic rotation velocity, $v_{\rm rot}$ of SP, we build its velocity field and rotation curve as a function of distance from the galaxy center. The VLT/MUSE spectral cube is cross-correlated, taking into account the spectral variance, with an [\ion{O}{ii}] template spectrum (see Appendix~\ref{sec:A1} for further details). The resulting velocity field is rotated by a position angle value estimated such that the elliptical fit of the galaxy is horizontal and the velocity map is collapsed along the major axis of that ellipse (see the left and middle panels of Figure \ref{velmap}). We measure, as shown in the right panel of Figure \ref{velmap}, a value of $v_{\rm rot}$ of $(94.5 \pm 5.3)$ km/s at a radial distance $r$ of $\sim 1.4\arcsec$, which corresponds to $\sim 8.7$ kpc at $z=0.511$. The actual value of $v_{\rm rot}$ depends on the inclination angle of SP, $i$, which appears to be nearly face-on (see Figure \ref{sdss1029_fig}). We estimate the value of the inclination angle of SP from the light model of the disc obtained with a S\'ersic profile in Section \ref{sec:LLM}. From the value of the minor-to-major axis ratio, $q_{\rm L}$ (see Table \ref{tab:light_models} and Figure \ref{lightmodels}), we estimate $i=35^{+6}_{-3}$ degrees. Given the challenge of measuring accurately the inclination angle of nearly face-on spiral galaxies and the residuals from our light model, we adopt a conservative uncertainty of $\pm10$ degrees for the value of $i$ \citep[see also e.g.,][]{DiTeodoro2018}.
Thus, from Equation \ref{Eq:Mdyn}, we obtain a measurement of the total mass of SP of $\sim (5.4^{+4.6}_{-1.9}) \times 10^{10} M_{\odot}$ at $1.4\arcsec$ ($\sim 8.7$ kpc at $z=0.511$) from the center of the galaxy, in agreement with those found by other studies at similar redshifts \citep[e.g.,][]{Moran2007}. 

We then measure the total mass of SP within a spherical radius, $r$, which can be expressed for a circular dPIE mass density profile, with a vanishing core radius, as 
\begin{equation}
M_{\rm lensing}^{\rm dPIE}(r) = \frac{2 \sigma_0^2}{G} ~ r_{\rm cut} ~ \arctan \left(\frac{r}{r_{\rm cut}}\right).
\label{Eq:MLensing_dPIE}
\end{equation} 
From the full MCMC chains, we estimate $M_{\rm Lenstool}^{\rm dPIE}(\le1.4\arcsec) = (2.4^{+0.4}_{-0.4}) \times 10^{11} M_{\odot}$ and $M_{\GLEE}^{\rm dPIE}(\le1.4\arcsec) = (1.1^{+0.1}_{-0.1}) \times 10^{11} M_{\odot}$. We thus find that the total mass of SP, derived from our \GLEE\ extended lens model, is in good agreement, within the uncertainties, with that derived from the VLT/MUSE [\ion{O}{ii}] gas kinematics analysis.
This result further showcases the importance of including the surface-brightness distribution of strongly-lensed sources as an observable in cluster lens models, allowing us to accurately probe the total masses, not only of cluster galaxies \citep{Grillo2008, Monna2015, Monna2017, Galan2024}, but also of structures along the line of sight lying angularly close to extended multiple images.

\section{Conclusions} \label{sec:conclu}
In this work, we have modeled the extended surface-brightness distribution of a strongly-lensed source, for the first time in a full strong-lensing model of a galaxy cluster, overcoming both modeling and computational challenges (allocating $\sim 100$ GB of RAM for several months). To do so, we have considered as testbed the galaxy cluster SDSS~J1029$+$2623 \citep[$z=0.588$,][]{Inada2006, Oguri2008}, and the strongly-lensed QSO host galaxy, observed in the HST/F160W filter with a high-signal-to-noise ratio (see Figure \ref{sdss1029_fig}). 

We have exploited high-resolution HST imaging and extensive VLT/MUSE spectroscopy of the galaxy cluster core \citep{Acebron2022} to construct a new, extended strong lensing model, in a full multi-plane formalism, by leveraging the unique capabilities of the \GLEE\ software \citep{Suyu2010, Suyu2012}.
The total mass distribution of SDSS~J1029$+$2623 has been modeled following \citet{Acebron2022}, where two line-of-sight galaxies angularly-close to the QSO multiple images, the foreground spiral ($z=0.511$) and the background galaxy GX ($z=0.674$), have been considered instead at their measured redshifts (see Figure \ref{sdss1029_fig}).
Moving beyond typical strong-lensing modeling techniques of galaxy clusters, we have included as observables both the positions of 26 \emph{pointlike} multiple images from seven background sources, spanning the wide redshift range from $z=1.02$ to $z=5.06$, and the extended surface-brightness distribution of the strongly-lensed QSO host galaxy over $\sim78000$ HST pixels in the WFC3/F160W band (with a $0\arcsec.03$ pixel scale). 
We have also taken into account the light contamination from seven angularly-close objects by modeling their surface-brightness distribution over $\sim9300$ HST pixels (see Figures \ref{ErrMasks} and \ref{lightmodels}). 

We have shown that our extended lens model reproduces well both the observed surface brightness and morphology of the QSO host galaxy in the HST F160W band (see Figure \ref{arcmodel}).
The reconstructed surface-brightness distribution of the source exhibits a smooth and compact component and is best modeled with a single S\'ersic profile (see Figure \ref{sourceQSO}). We have characterized the intrinsic properties of the QSO host galaxy at $z=2.1992$, inferring a F160W magnitude of $23.3\pm 0.1$ and a half-light radius of (2.39 $\pm$ 0.03) kpc. With this work, we have thus highlighted the potential of leveraging cluster-lensed QSO for extending current studies, both with galaxy-scale lensed and non-lensed QSOs systems, of the correlation between the supermassive black-hole mass and host-galaxy properties, to $\sim 1$ magnitude fainter \citep[e.g.,][]{Peng2006, Ding2017b}. 

While the total cluster mass distributions obtained from our \emph{position-based} or extended strong-lensing models are very similar (see Figure \ref{kappa}), the increased number of observables allows for the reduction of model degeneracies and results in a more precise measurement of the mass parameters, as clearly illustrated in Figure \ref{SP_GX}. We have also obtained an accurate determination of the total mass parameters associated with halos lying angularly close to the extended arc (see Figure \ref{SP_GX}). Specifically, we have found that the total mass of the foreground spiral galaxy SP, measured from our extended model, is in agreement, within the uncertainties, with that estimated from the [\ion{O}{ii}] gas kinematics analysis, enabled by our VLT/MUSE observations. 
We have also shown that it is possible to infer the lensing contribution of a very small perturber (the hidden galaxy, GX) in a “blind way”, as the low-mass halo is barely visible in the HST imaging, due to the close and bright QSO multiple image C. 

With the first light of cutting-edge facilities such as the JWST, Euclid and Roman telescopes, and Rubin Observatory's Legacy Survey of Space and Time, this decade is set to witness a golden era for strong gravitational lensing with galaxy clusters.
This work paves the way for a new generation of cluster strong-lensing models, where all available lensing observables are incorporated as model constraints. As such, promising prospects for time-delay cosmography, QSO host-galaxy properties, and mass substructures studies in lens galaxy clusters are within reach.


\begin{acknowledgements}
We kindly thank the referee for the comments received, which have helped improving the manuscript.
This work is based in large part on data collected at ESO VLT (prog. ID 0102.A-0642(A)) and NASA HST. 
AA has received funding from the European Union’s Horizon 2020 research and innovation program under the Marie Skłodowska-Curie grant agreement No 101024195 — ROSEAU and acknowledges financial support through the Beatriz Galindo program and the project PID2022-138896NB-C51 (MCIU/AEI/MINECO/FEDER, UE), Ministerio de Ciencia, Investigaci\'on y Universidades.
We acknowledge financial support through grants PRIN-MIUR 2017WSCC32 and 2020SKSTHZ.
SHS, SE, and HW thank the Max Planck Society for support through the Max Planck Research Group and the Max Planck Fellowship for SHS.
SS has received funding from the European Union’s Horizon 2022 research and innovation program under the Marie Skłodowska-Curie grant agreement No 101105167 - FASTIDIoUS.
\end{acknowledgements}

%

\facilities{HST(ACS,WFC3), VLT(MUSE)}






\appendix
\restartappendixnumbering
\label{appendix_A0}
\section{Extended reconstruction of the strongly-lensed QSO host galaxy from a \emph{position-based} lens model} \label{sec:A0}
\begin{figure*}[ht!]
\centering
\includegraphics[width=\linewidth]{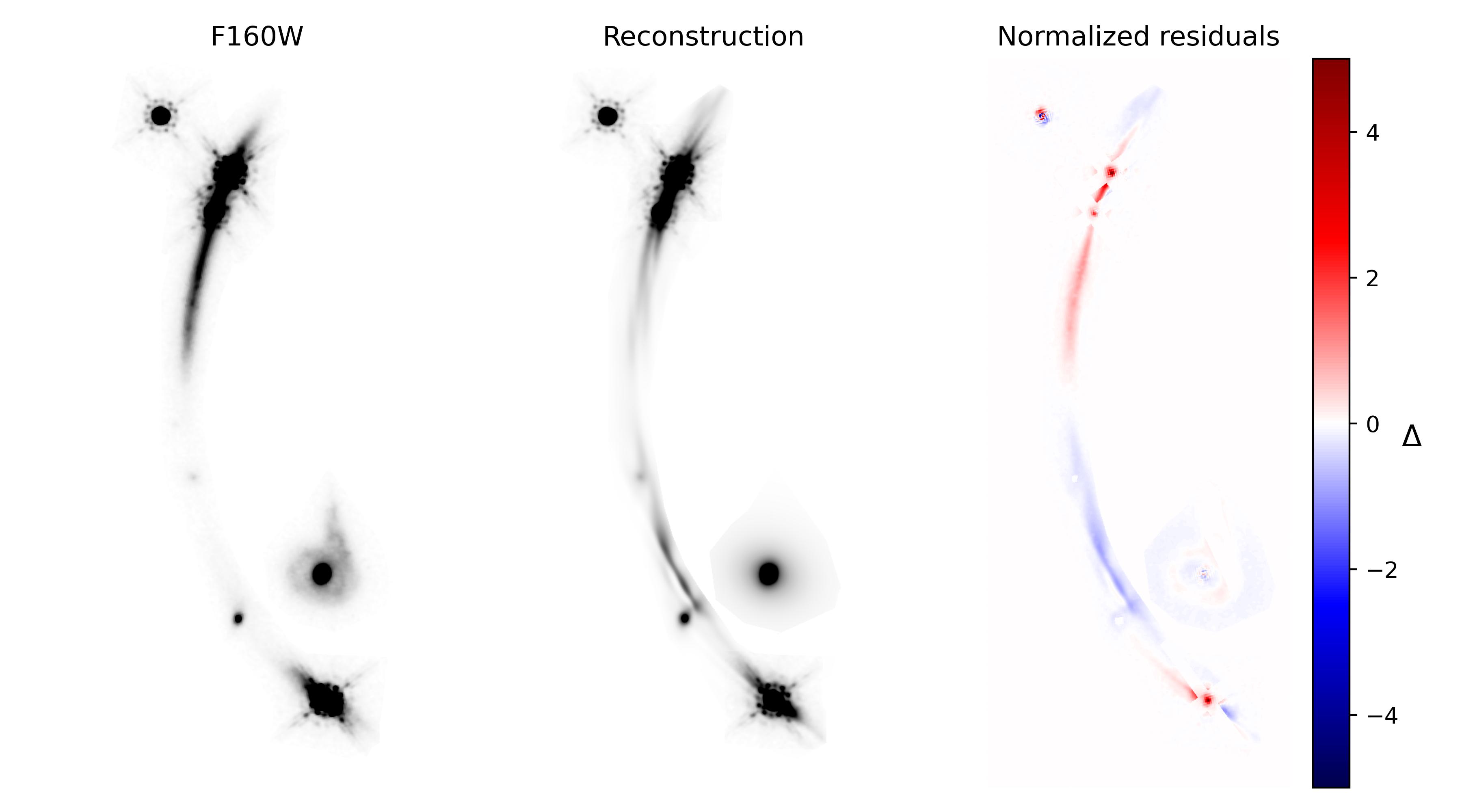} 
\caption{Reconstructed surface-brightness distribution of the QSO host galaxy at $z=2.1992$ from the \emph{position-based} GLEE lens model (see Sections~\ref{sec:PLMod} and \ref{sec:results}), where the surface-brightness distribution is not considered as an observable in the lens model. The same panels, field of view and orientation as in Figure~\ref{arcmodel} are shown. 
} 
\label{extreconstruction}
\end{figure*}

Appendix~\ref{sec:A0} presents the extended reconstruction of the strongly-lensed QSO host galaxy from the GLEE \emph{position-based} best-fit lens model, which only considers as observables the positions of the \emph{pointlike} multiple images (see Section~\ref{sec:PLMod}).
The results are shown in Figure~\ref{extreconstruction}, highlighting the poor reconstruction compared to that from our extended model shown in Figure~\ref{arcmodel}. In particular, we note that the values of the normalized residuals around the galaxy G1 reach values in the reconstructed surface-brightness distribution of a factor $>100$ worse than in our extended model (see Figure~\ref{arcmodel}). This is due to the fact that the total mass contribution of SP is more accurately constrained thanks to the direct modeling of the extended surface-brightness distribution of the QSO host galaxy (see Section~\ref{sec:TMD}). 

\label{appendix_A}
\section{Gas kinematic analysis of the foreground spiral galaxy} \label{sec:A1}
\begin{figure*}[ht!]
\centering
\includegraphics[width=\linewidth]{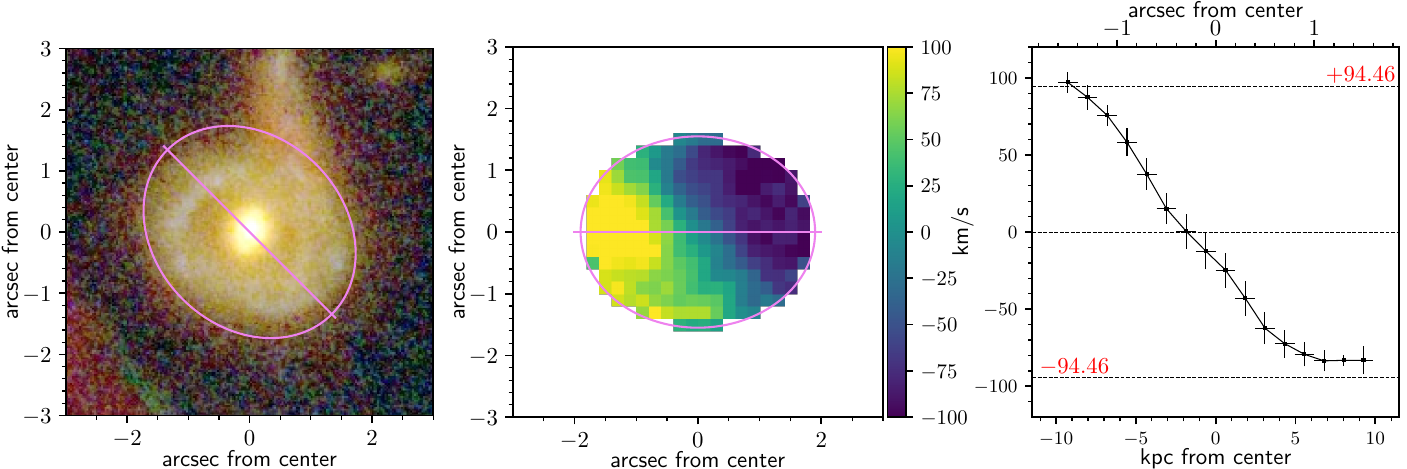} 
\caption{Results from the [\ion{O}{ii}] gas kinematic analysis of the foreground spiral galaxy, SP. \textbf{Left:} HST color cutout centered on SP. \textbf{Middle:} Rest-frame velocity field of SP derived from the [\ion{O}{ii}] emission line, and using a systemic redshift of 0.511 for the center of the galaxy. \textbf{Right:} Derived rotation velocities projected along the kinematic major axis, indicated with the pink solid line in the velocity map.} 
\label{velmap}
\end{figure*}

Appendix~\ref{sec:A1} presents the [\ion{O}{ii}] gas kinematic analysis of the foreground spiral galaxy (SP, $z=0.511$) included in our multi-plane, extended strong-lensing model. The [\ion{O}{ii}] template used is obtained by stacking the [\ion{O}{ii}] emission galaxy rest-framed spectra with $z \sim 0.5-1.1$ \citep[taken from][]{Bergamini2019, Bergamini2021, Vanzella2021}. The velocity map is then built as $c\left(\frac{z_{ij}-z_0}{1+z_{ij}}\right)$, where $z_{ij}$ is the best-fitting redshift value for the pixel $i, j$ and $z_0$ is the measured systemic redshift of $z=0.511$. The methodology will be presented in detail in Angora et al. (in prep).
In Figure \ref{velmap}, we show the velocity field and derived projected rotation curve, yielding a measurement of the asymptotic rotation velocity of SP of $94.5 \pm 5.3$ km/s.

\bibliography{sample631}{}
\bibliographystyle{aasjournal}



\end{document}